\documentclass[onecolumn,prl]{revtex4}
\usepackage{graphicx,psfrag,dsfont}
\usepackage{slashed,amssymb,color}
\def\bc{\begin{center}}
\def\ec{\end{center}}

\def\beq{\begin{equation}}
\def\eeq{\end{equation}}

\def\u{\uparrow}

\newcommand{\nn}{\nonumber}

\begin{document}
 
\title{ 
Electron pairing with gapless excitations in mixed double layers 
}
\author{Andreas Sinner$^{1}$, Yurii E. Lozovik$^{3,4}$, and Klaus Ziegler$^{1}$}
\affiliation{
\mbox{$^{1}$ Institut f\"ur Physik, Universit\"at Augsburg, D-86135 Augsburg, Germany}\\
\mbox{$^{3}$ Institute of Spectroscopy, Russian Academy of Sciences,142190 Troitsk, Moscow, Russia}\\
\mbox{$^{4}$ Moscow Institute of Electronics and Mathematics, National Research University}\\
\mbox{Higher School of Economics, 101000 Moscow, Russia}
}

\begin{abstract}
We study the interlayer pairing states in layered systems of two different 2d electronic subsystems, one with relativistic linear and the other with non-relativistic parabolic spectrum. 
The complex order parameter of the paired state has a two component structure. 
We investigate the pairing state formation on the mean-field level, determine the critical interaction strength and evaluate the effective potential.
The anisotropic three-band spectrum of quasiparticles depends explicitly on the phase difference of the order parameter components,
rotates in momentum space as it changes, and exhibits the strong band deformation due to the pairing.
The pairing leads to the fusion and hybridization of initially decoupled bands.
The quasiparticle spectrum has the shape of deformed Dirac cones in the vicinity of the two touching points between neighboring bands. 
The density of states exhibits a number of specific features due to band deformation, such as a van Hove singularity.
\end{abstract}

\maketitle

\section{Introduction} 

Often, the layered electronic systems disclose physical phenomena, which are neither present in a single 2d layer, nor in isotropic 3d systems.
The role of the layered structures is or might be important for understanding physical phenomena as different as the 
formation of interlayer exciton condensates in semiconducting devices~\cite{Lozovik1976,Lozovik2011,Lozovik2012,Lozovik2018,MacDonald2004}, the interplay of excitonic superfluidity with
unconventional fractional quantum Hall states in graphene bi- and double layers in external magnetic fields~\cite{Dean2017,Dean2019,Kim2019}, 
the complex of problems with the high-T$_c$ superconductivity~\cite{bednorz86,kettemann92}, 
the crossover between adjacent superconducting and insulating states in magic angle twisted graphene bilayers~\cite{cao18a,cao18b}, 
the anomalous giant magneto-resistance and superconductivity in graphene~\cite{song18,saito16,murata19}, the modeling of the Hubbard physics by moir\'{e} excitons in WSe2/WS2 heterobilayer~\cite{MacDonald2020},
and transition metal dichalcogenide (TMDC) multilayers~\cite{geim13,Fogler2014}.
Recently we proposed a set of pairing states that can emerge in a layered system of two graphene layers due to repulsion between electrons from opposite layers~\cite{SLZ20}. 
We also pointed out a duality between electron-electron and electron-hole condensates occurring in layered graphene devices.

Layered structures consisting of two 2d electronic systems, one with linear relativistic (Dirac) dispersion and another with non-relativistic parabolic (conventional) dispersion
have received some attention in the past. 
Suggestions were made that sandwiches of graphene and gallium arsenide layers can host similarly inhomogeneous excitonic electron-hole condensates~\cite{Polini2012,Linh2018,Phuong2019,Gamucci2015}. 
Interacting 2d Dirac fermions and 2d non-relativistic electrons might coexist on the surface of a 3d topological insulator (e.g. in $\rm Bi^{}_2Se^{}_3$)~\cite{Bianchi2010,Madhavan2013,Dou2014}. 
For the larger part though, the theoretical works were restricted to the studies of the plasmon spectrum~\cite{Polini2012,Jain2014}. 
In particular, the setup of Ref.~\cite{Jain2014} modeled both electron species as confined to two spatially separated layers. 
The fine-tuning of the strength of the repulsive interspecies interaction was interpreted as the variation of the spatial separation between the layers. 
Such bilayer systems may represent environments in which electron pairing between different species occurs due to Coulomb repulsion,  
in analogy to the case of graphene bilayer considered by us in Ref.~\cite{SLZ20}. 
In this paper we study such interlayer paired states. 
We find an anisotropic quasiparticle dispersion $E(q_x,q_y)$, although the dispersion without pairing is isotropic. 
The existence of anisotropic quasiparticle dispersion is quite common in many branches of physics,
often associated with the phenomena like birefringence~\cite{Bartholin,Landafshitz,BornWolf} 
or electronic nematicity~\cite{Fradkin2010,Fernandes2014,Fernandes2019,cao20}.
The order parameter of the paired state has two complex components. 
Among other features, touching points between neighboring bands form in the spectrum of the mean-field Hamiltonian. 
Close to those points the quasiparticle spectrum has the shape of deformed Dirac cones. Because of this, the density of states exhibits a van Hove singularity.

\begin{figure}[t]
 \includegraphics[width=9cm]{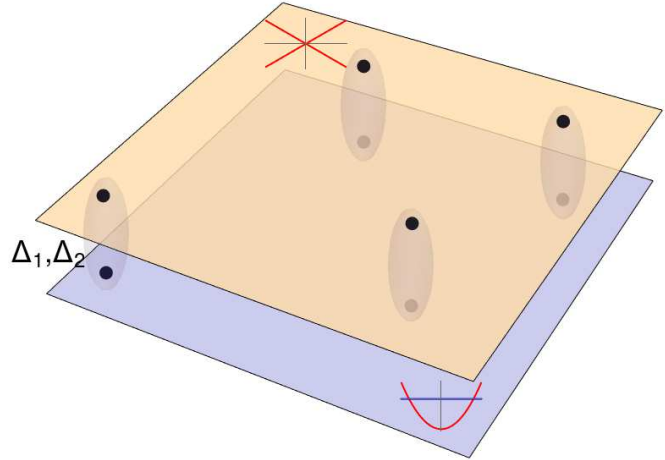}
\caption{
The schematic representation of the considered system. The upper layer is supposed to be populated by the Dirac electrons with linear spectrum, 
the lower layer by conventional electrons with parabolic dispersion. The electrons from both layers (black dots) form two-component pairing order parameter emphasized by the enveloping clouds.
}
\label{fig:Fig1}
\end{figure}

\section{Microscopic model and the mean-field approximation}

The proposed system is composed of a charge neutral graphene layer and a layer of a layer hosting the conventional 2d electron gas with the Fermi level risen into the band.
By bringing them close to each other both 2d electron gases feel the mutual Coulomb repulsion, which results in a formation of an ordered interlayer state of paired particles as shown in Fig.~\ref{fig:Fig1}. 
We describe this system by a microscopic second quantized Hamiltonian which neglects the spin degrees of freedom and models the particles with relativistic spectrum by a single-cone Dirac Hamiltonian.
The model Hamiltonian reads
\begin{equation}
\label{eq:McH}
{\rm H} = \psi^\dag\cdot[- iv\nabla^{}_1\sigma^{}_1- iv\nabla^{}_2\sigma^{}_2 + \Delta^{}_{\rm D}\sigma^{}_3]\psi + \varphi^\dag\cdot[ -\frac{\nabla^2}{2 m} - \mu]\varphi + 
I^{}_g[\psi,\varphi]
\end{equation}
where $\psi^{}=(\psi^{}_1,\psi^{}_2)$ and $\varphi$ are the annihilation (and the corresponding creation) operators acting in the layers with the relativistic and non-relativistic spectrum respectively;
$\nabla^{}_{1,2}$ are spatial derivatives in respective direction and $\nabla^2=\nabla^2_1+\nabla^2_2$;
$v$ is the Fermi velocity of the Dirac electron, $\mu$ the chemical potential of the conventional electron gas, $m$ the band mass of conventional electrons, and  
$\sigma^{}_{1,2,3}$ are the Pauli matrices in usual representation, which act on the Dirac space. 
$\Delta^{}_{\rm D}$ denotes the Dirac mass, which might be attributed to the intrinsic spin-orbit coupling. 
Its  sign is not fixed, i.e. $\Delta^{}_{\rm D}$ can be negative. 
We do not specify the interlayer Coulomb interaction term $I^{}_g[\psi,\varphi]$ here, which might be very general and only has to be given in terms of particle densities of both species. 
Finally, the dot operator denotes the integration in the 2-dimensional position space.
In both layers, the intralayer Coulomb interaction is supposed to be strongly suppressed due to screening and therefore negligible in first approximation.
For conventional electrons the screening is due to the finite density of states at the Fermi surface, while in the case of Dirac electrons the screening is due to the Schwinger 
particle-hole production at the Dirac point, cf. Ref.~\cite{SLZ20} and references therein, 
or by disorder or by thermal and electrostatic fluctuations~\cite{Liu2009}.
In order to capture the effects qualitatively, it is often sufficient to approximate the extended interlayer interaction by a simple contact interaction~\cite{Stoof2011,Berman2019,SLZ20,MacDonald2020,Fogler2014}.
Experimentally, the strength of the interalyer interaction is amenable 
by changing the dielectric material between the layers~\cite{Kim2019,Dean2017,Dean2019,MacDonald2020,Fogler2014}.

In the mean-field approximation the electron pairing appears in form of the two-component order parameter with spinor structure. 
The anticipated mean-field Hamiltonian reads
\begin{equation}
\label{eq:MFH}
{\rm H}^{}_{\rm MF} = 
\left(
\begin{array}{c}
\psi^\dag_1 \\
\psi^\dag_2 \\
\varphi^\dag
\end{array}
\right)^{\rm T}_q \cdot
\left(
\begin{array}{ccc}
         \Delta^{}_D              &   vq e^{i\phi}        &  \Delta^{}_1 e^{i\chi^{}_1}  \\  
   vq e^{-i\phi}     &          - \Delta^{}_D                &   \Delta^{}_2 e^{i\chi^{}_2} \\
 \Delta^{}_1e^{-i\chi^{}_1} & \Delta^{}_2 e^{-i\chi^{}_2} &  \xi^{}_q        
\end{array}
\right)
\left(
\begin{array}{c}
\psi^{}_1 \\
\psi^{}_2 \\
\varphi
\end{array}
\right)^{}_q,
\end{equation}
where  $q=\sqrt{q^2_x+q^2_y}$, 
$\phi={\rm atan}\left[\frac{q^{}_y}{q^{}_x}\right]$, $q^{}_x$ and $q^{}_y$ being the components of the momentum vector, 
and $\xi^{}_q~=~\frac{q^2-q^2_F}{2m}$ with the Fermi momentum related to the chemical potential  $\mu = \frac{q^2_F}{2m}$.
$\Delta^{}_1$ and $\Delta^{}_2$ are positive amplitudes and $\chi^{}_1$ and $\chi^{}_2$ the global phases 
of the two-component complex order parameter corresponding to each respective sublattice. 
The Hamiltonian~(\ref{eq:MFH}) is invariant under a simultaneous global U(1) transformation of both order parameters.
It is rather generic and does not rely on any particular interlayer interaction term. 
As an example, in Appendix~1 we show how the order parameter of Hamiltonian~(\ref{eq:MFH}) emerges from the simplest contact density-density interaction.
The global rotation $\psi^{}_{j} = e^{i\chi^{}_{j}}\psi^{}_{j}$, $j=1,2$ and $\psi^\dag_{j} = e^{-i\chi^{}_{j}}\psi^\dag_{j}$ in the Hamiltonian Eq.~(\ref{eq:MFH}) 
leaves the diagonal elements of the kernel matrix unchanged, eliminates the phase of the order parameters and shifts the phase of the complex momentum in the Dirac layer as
$\phi\to\phi+\chi^{}_2-\chi^{}_1$. Therefore, the mean-field Hamiltonian depends only on the total phase $\phi+\chi^{}_2-\chi^{}_1$
and the phase difference $\chi^{}_2-\chi^{}_1$ rotates the momentum, and therefore the spectrum of the Hamiltonian. 

\begin{figure}[t]
\includegraphics[width=8.5cm]{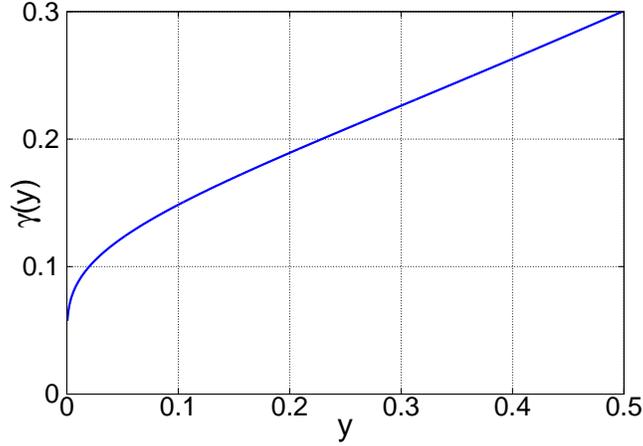}
\caption{
The critical interaction strength of pairing transition  as function of $y=mv/q^{}_F$, where 
$v$, $m$ and $q^{}_F$ represent the Fermi velocity of Dirac particles, as well as the band mass 
and the Fermi momentum of the conventional particles, respectively.
}
\label{fig:Fig2}
\end{figure}

\begin{figure}[t]
\includegraphics[width=4.75cm]{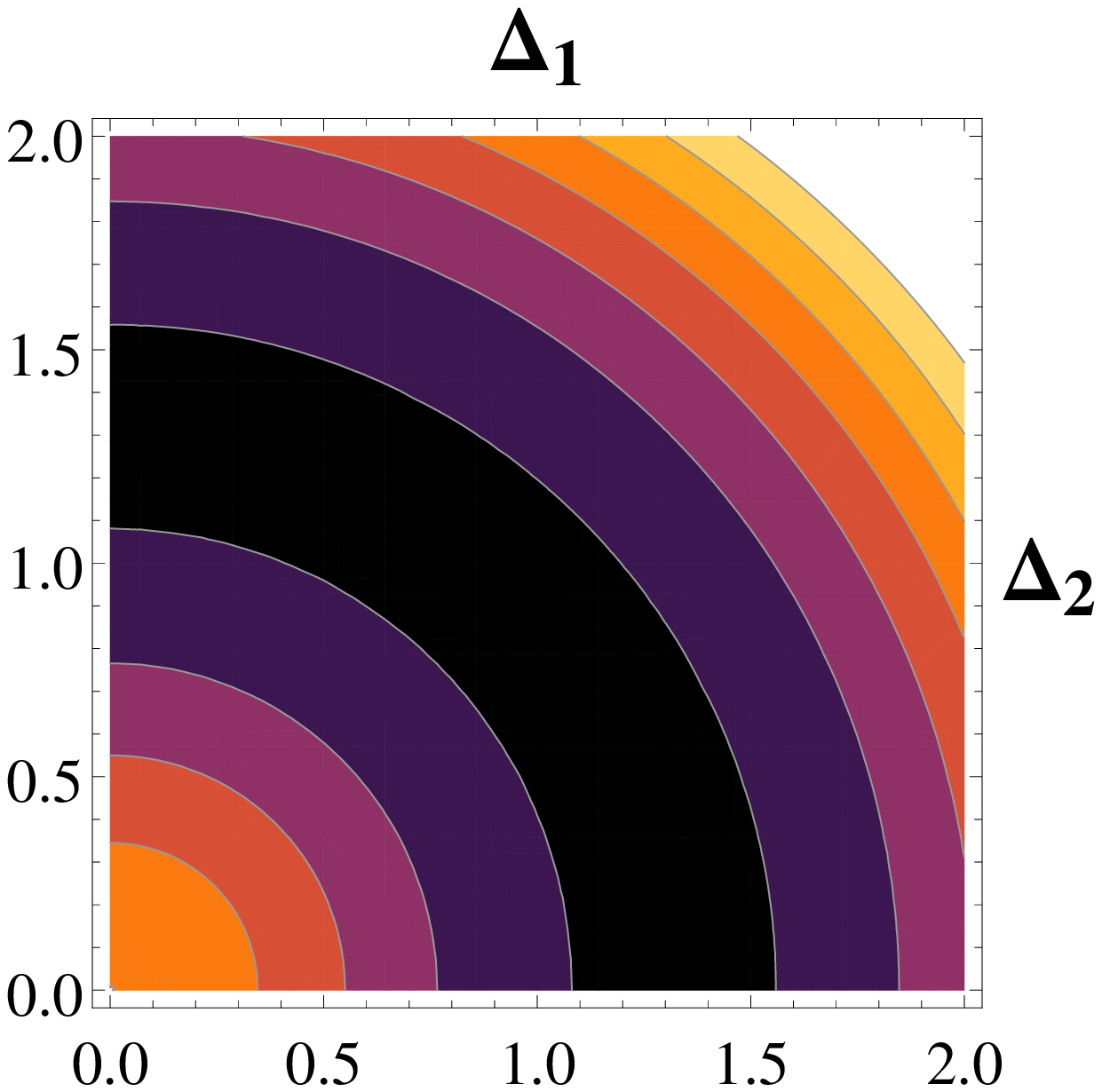}
\hspace{5mm}
\includegraphics[width=4.75cm]{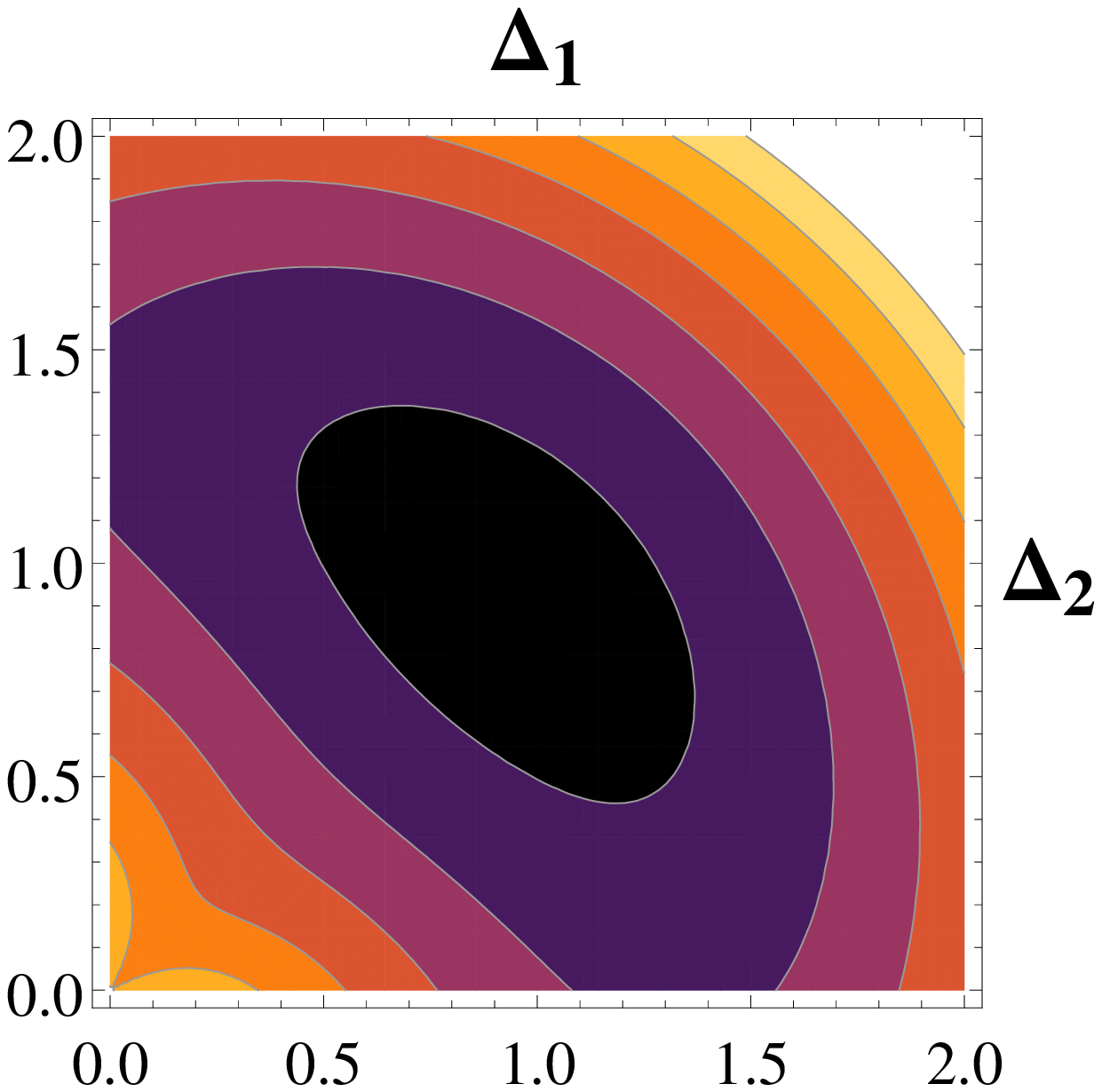}
\\
\includegraphics[width=4.75cm]{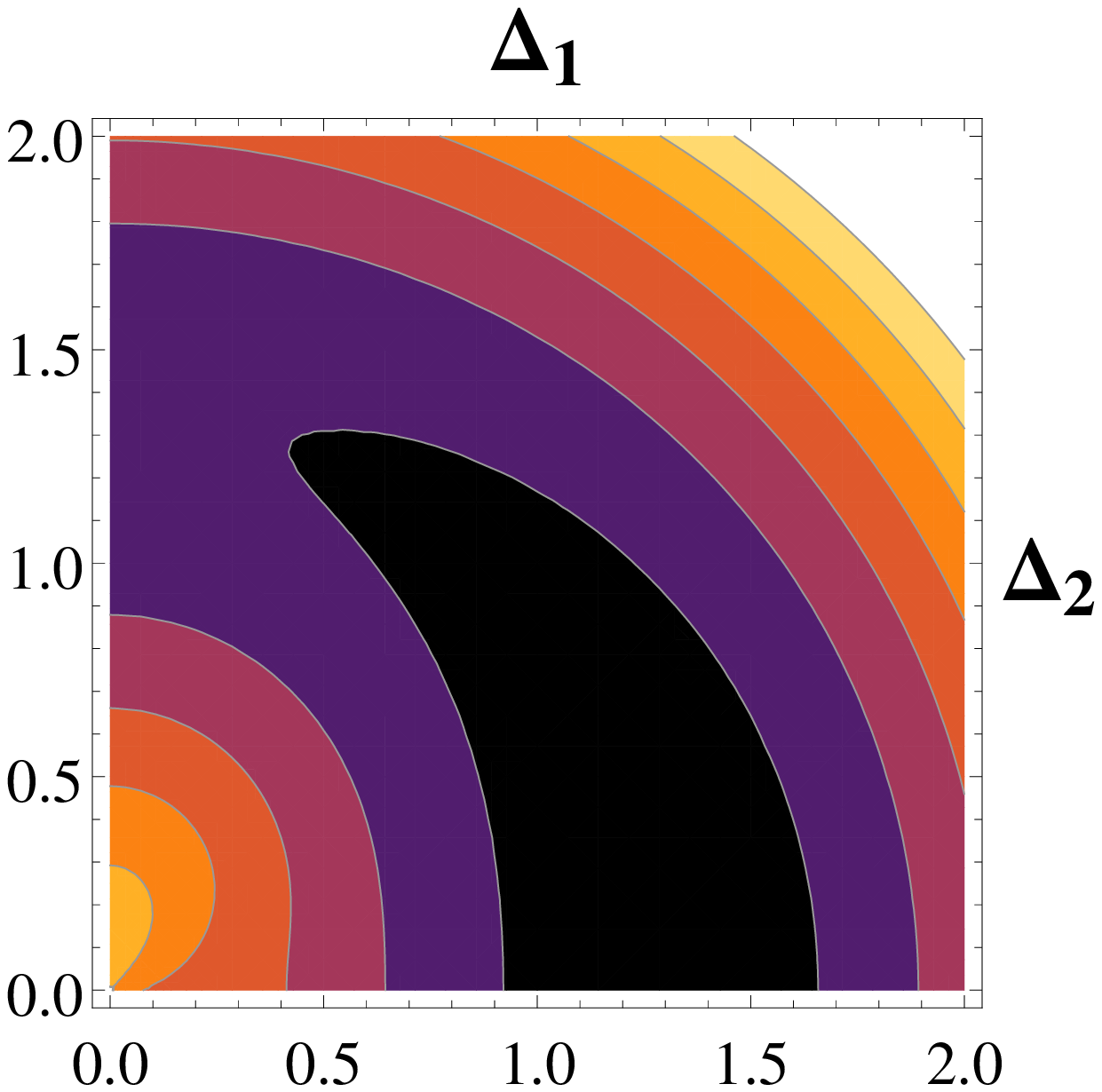}
\hspace{5mm}
\includegraphics[width=4.75cm]{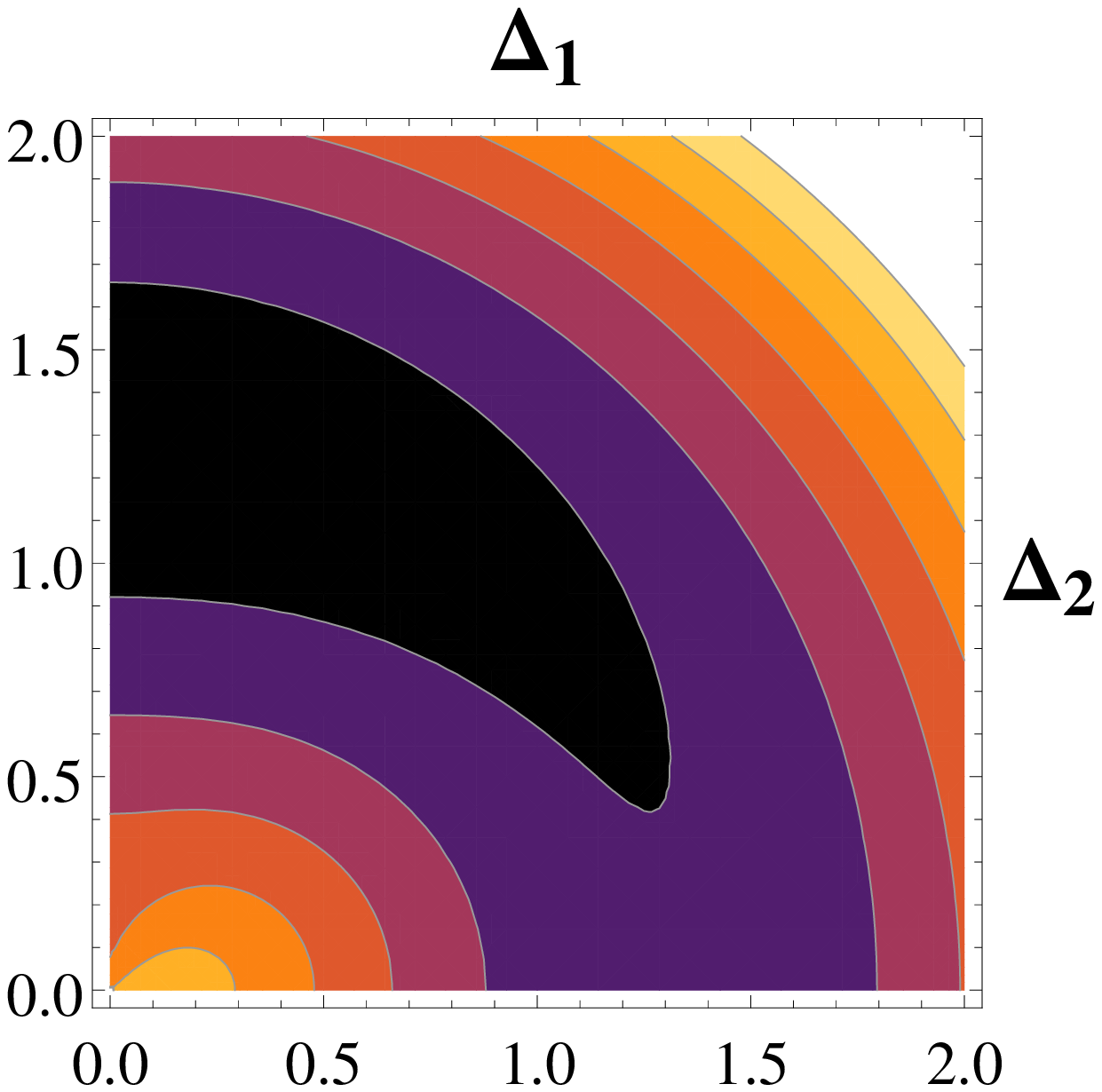}
\caption{
Four screenshots of the effective potential Eq.~(\ref{eq:fef}). 
Upper row left panel: The case with zero Dirac mass $\Delta^{}_{\rm D}=0$ and $v\Lambda/\mu=0$, $\Lambda$ being the band width, with a degenerate minimum.
Right panel: The case with zero Dirac mass $\Delta^{}_{\rm D}=0$ and $v\Lambda/\mu=0.5$. 
The ground state lies in the local minima along the line $\Delta^{}_1=\Delta^{}_2$.
Bottom row: The effective potential for positive and negative Dirac mass $\Delta^{}_{\rm D}$ and $v\Lambda/\mu=0$ respectively, 
with minima placed along one of the axis.
}
\label{fig:Fig3}
\end{figure}

\section{Generation of the pairing order parameter}

The generation of the order parameter is captured by the zero temperature effective potential of the paired phase, which for the contact interaction is defined as 
\begin{equation}
\label{eq:fef}
{\cal F}^{}_{MF} = \frac{1}{2g}\left(\Delta^2_1+\Delta^2_2\right) - \int\frac{dq^{}_0}{2\pi}~\int^\Lambda\frac{d^2q}{(2\pi)^2}~\log\det 
\left[
iq^{}_0 + {\rm H}^{}_{\rm MF}
\right],
\end{equation}
where $g$ is the interaction strength and the integration over the imaginary frequency $q^{}_0$ stretches from $-\infty$ to $+\infty$ and the radial 
momentum integration stretches from $0$ to the upper cutoff $\Lambda$. 
The variation of this functional with respect to each of $\Delta$'s provides us with the system of mean-field equations, from which  
the critical interaction strength for the pairing (i.e. for $\Delta^{}_{\rm D} = 0$, $\Delta^{}_{1} = \Delta^{}_{2} = 0$) follows. 
The critical interaction strength condition reads 
\begin{equation}
\label{eq:CritCond}
1 = 2g^{}_c \int\frac{dq^{}_0}{2\pi}~\int^\Lambda\frac{d^2q}{(2\pi)^2}~\frac{q^2_0}{(q^2_0+q^2v^2)(q^2_0+\xi^2_q)}.
\end{equation}
The evaluation of the integral on the right hand side is presented in the Appendix~2. The result of the integration is
\begin{eqnarray}
\nn
\frac{4\pi}{mg^{}_c} &=& \left[1-\frac{y}{\sqrt{1+y^2}}\right]\left(\log\left[\frac{\lambda+y-\sqrt{1+y^2}}{1+y-\sqrt{1+y^2}}\right] - \log\left[\frac{\sqrt{1+y^2}-y+1}{\sqrt{1+y^2}-y}\right]\right)   \\
\label{eq:critg}
&+& \left[1+\frac{y}{\sqrt{1+y^2}}\right]\left(\log\left[\frac{1+y+\sqrt{1+y^2}}{y+\sqrt{1+y^2}}\right] - \log\left[\frac{y+\sqrt{1+y^2}-1}{y+\sqrt{1+y^2}}\right]\right),
\end{eqnarray}
where $y=mv/q^{}_F$, and $\lambda=\Lambda/q^{}_F$. The inverse of the left hand side ($\sim mg^{}_c$) is plotted in Fig.~\ref{fig:Fig2} for $\lambda=10$.
The critical interaction strength turns out to be a universal function of the parameter $y=mv/q^{}_F$, e.g. the ratio of the two Fermi velocities $v$ and $q^{}_F/m$. 
In Fig.~\ref{fig:Fig1} we plot the dimensionless interaction strength $\gamma = gm/4\pi$ as a function of the parameter $y=mv/q^{}_F$.
It represents a monotonously increasing function which exists for all values of $y$. 
It approaches zero for $y\to0$, i.e. for $mv\ll q^{}_F$, which suggests that the chemical potential $\mu$ in the conventional layer is a fine-tuning parameter.

A persisting challenge of graphene physics concerns the question whether the intralayer Coulomb interaction is strong enough to open a gap in the spectrum of Dirac particles. 
It is therefore important to estimate the competition between the two tendencies. 
The actual quantity, which measures the dimensionless Coulomb interaction strength is the effective fine structure constant $\alpha=e^2/v$, 
which takes in the suspended graphene the value $\alpha\sim2.17$. The size of the critical value of gap opening $\alpha_c$ is somewhat
arguable and spreads in the literature in a range $\alpha_c\sim$1-10, cf. Ref.~\cite{Drut2008,Wang2012,Ulybyshev2013,Popovici2013,Kanoda2021} and references therein. Larger values seem to be more realistic, given the fact that no experiments have ever observed a gap opening in graphene at the Dirac point~\cite{Kanoda2021}. 
For instance, the computations performed by Wang and Liu in Ref.~\cite{Wang2012}, based on Dyson-Schwinger self-consistent approach, place the critical value of gap opening into the window 
$3.2\leqslant\alpha_c\leqslant 3.3$, which is much larger than the value of the free standing graphene. 
Thus, the intralayer Coulomb interaction cannot open a band gap in the free standing graphene monolayer. 
The situation is quite different for the interlayer interaction. This can be tuned over a large range by 
changing the dielectric parameter as well as by changing  the interlayer distance~\cite{Kim2019,Dean2017,Dean2019,MacDonald2020,Fogler2014}.

\section{The approximate effective potential} 

\begin{figure*}[t]
\includegraphics[width=4.5cm]{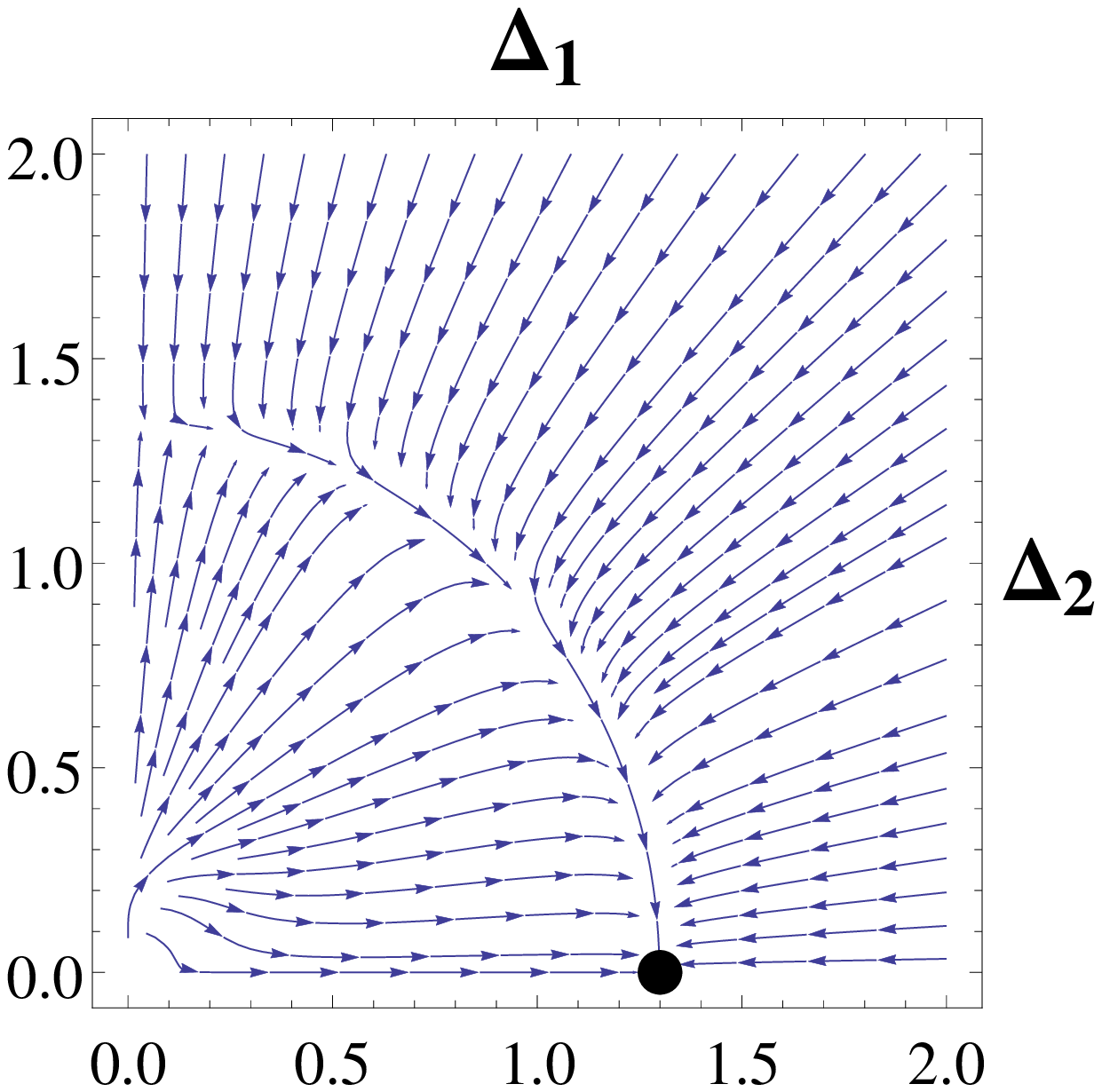}
\hspace{5mm}
\includegraphics[width=4.5cm]{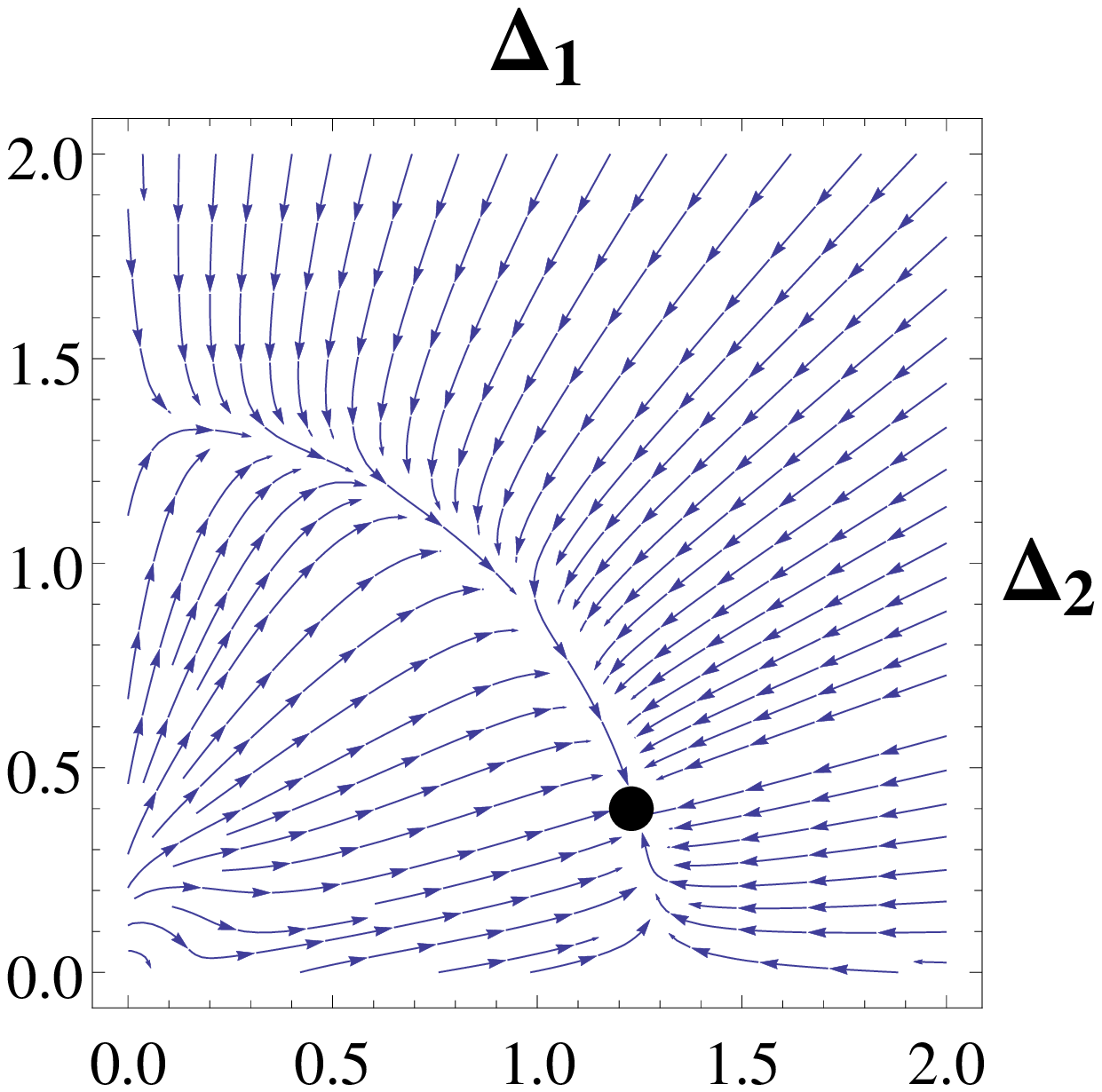}
\hspace{5mm}
\includegraphics[width=4.5cm]{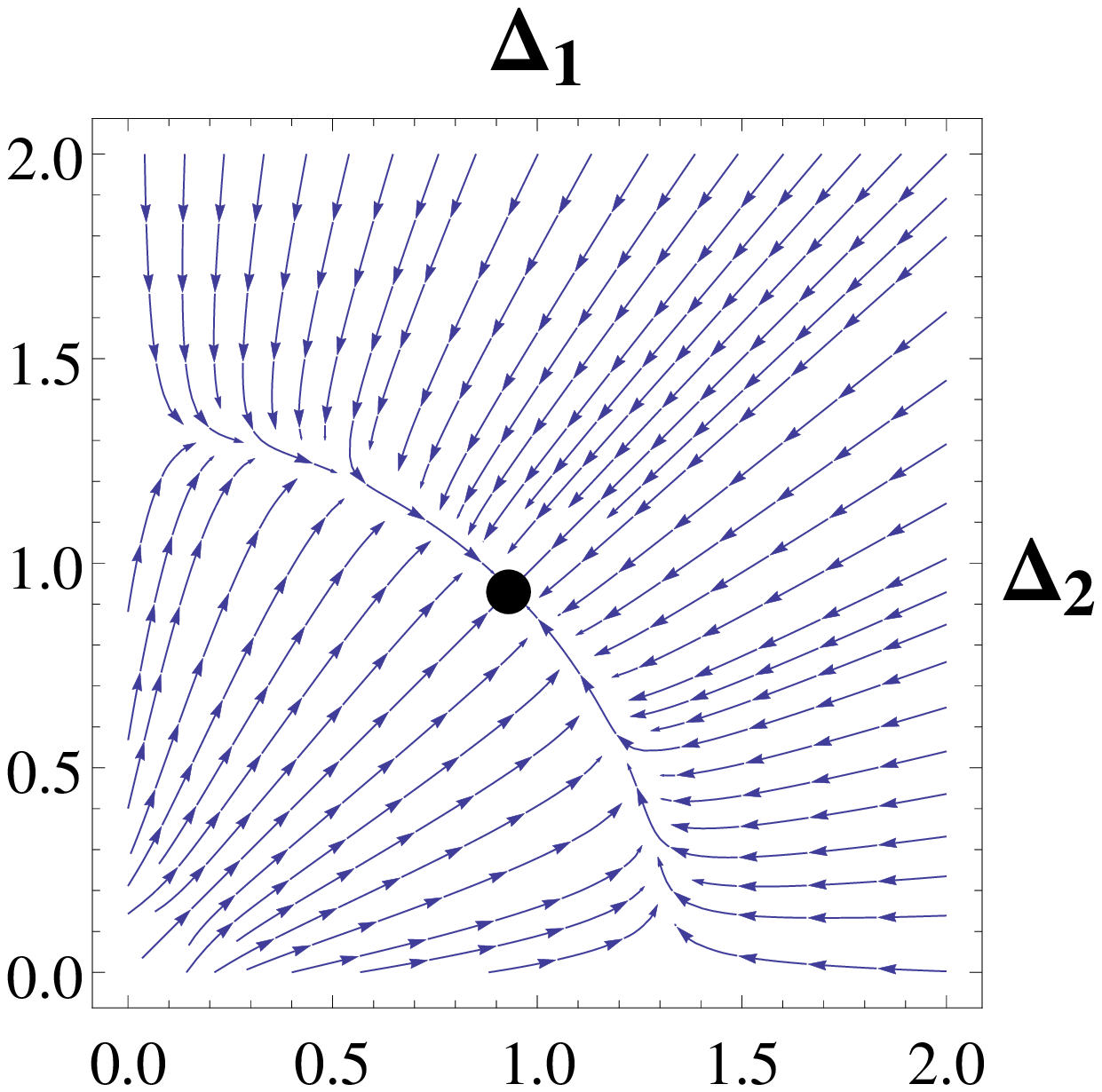}
\caption{
Gradient flow of the effective potential. Black dots denote the attractive points of the gradient flow, i.e. the solutions of the mean-field equations for order parameters.
The motion of attractive points is due to competition between model's parameters. 
Left: $v\Lambda/\mu=0$, $\Delta^{}_{\rm D}/\mu=0.25$; Middle: $v\Lambda/\mu=0.1$, $\Delta^{}_{\rm D}/\mu=0.25$;
Right: $v\Lambda/\mu=0.1$, $\Delta^{}_{\rm D}/\mu=0$. $\Lambda$ denotes the band width.
}
\label{fig:Fig4}
\end{figure*}

The approximate evaluation of the effective potential Eq.~(\ref{eq:fef}) is summarized in the Appendix~3.
Due to angular integration, the effective potential does not depend on the phase of the order parameters.
First term in Eq.~(\ref{eq:fef}) exhibits the order parameter symmetry $\Delta^{}_{1}\leftrightarrow\Delta^{}_{2}$.
On the other hand, the logarithm term in Eq.~(\ref{eq:fef}) has this symmetry only if we put  the Dirac mass ($\Delta^{}_{\rm D}=0$) to zero. 
If the Fermi velocity is put to zero ($v=0$) too, then the effective potential has a degenerate minimum, which is shown in Fig.~\ref{fig:Fig3}.
This regime can be also realized by an extreme rising of the Fermi energy in the conventional layer, i.e. $\mu\to\infty$.
There is a global U(1) symmetry when we simultaneously apply the same U(1) transformation to both pairing order parameters.
For zero Dirac mass $\Delta^{}_{\rm D}$ and non-zero $v$ there appears a unique minimum along the line $\Delta^{}_1=\Delta^{}_2$.
The invariance of the effective potential under the transformation $\Delta^{}_{1}\leftrightarrow\Delta^{}_{2}$ is broken by $\Delta^{}_{\rm D}$.
Depending on the sign of the Dirac mass, the minima in the potential appear along one of the axis, i.e. in this minimum one of the order parameters is fully suppressed. 
Finally, if both the Fermi velocity and the Dirac mass are finite, then there is a competition between both tendencies. 
To visualize this competition, it is more convenient to plot not the effective potential but rather its gradient flow, shown in Fig.~\ref{fig:Fig4}. 
Black dots in this plot denote the position of the attractive points in which the gradient of the effective potential vanish, i.e. 
these points represent the solutions of the variational equations for the order parameters. 

\section{Band crossing points in the spectrum of the Hamiltonian~(\ref{eq:MFH})}

The first insights into the spectral properties of the Hamiltonian~(\ref{eq:MFH}) are gained by inspection of the determinant of its kernel matrix. 
With $\Delta^{}_D=0$ it reads
\begin{eqnarray}
\label{eq:detGmf}
\det[{\rm H}^{}_{\rm MF}] &=& 2vq\Delta^{}_1\Delta^{}_2 \cos\left(\chi^{}_{2} - \chi^{}_{1} + \phi\right) -  v^2q^2\xi^{}_q ,
\end{eqnarray}
where again $q=\sqrt{q^2_x+q^2_y}$ and $\phi={\rm atan}\left[\frac{q^{}_y}{q^{}_x}\right]$. 
The determinant and therefore the spectrum depends on the combination $\chi^{}_{2} - \chi^{}_{1} + \phi$. This implies that the change of the phase difference 
of the order parameters $\chi^{}_{2} - \chi^{}_{1}$ is equivalent with the rotation of the momentum vector $q$ by this phase difference in the momentum space. 
The Dirac mass $\Delta^{}_D$ breaks the symmetry of both order parameter components $\Delta^{}_{1}\leftrightarrow\Delta^{}_{2}$, 
which is also reflected by the saddle-point solution, cf. Fig.~\ref{fig:Fig4}, and hence is likely to create the non-zero phase difference $\chi^{}_{2} - \chi^{}_{1}$. 

The determinant is a quartic polynomial in $q$. 
When $\mu=0$ the determinant is proportional to $q^4$ for $\Delta_1=\Delta_2=0$ and for $\Delta_1, \Delta_2 \neq 0$ proportional to $q$ for small momenta.
Since the determinant is the product of the energy eigenvalues, this behavior reflects a drastic change of the dispersion at small $q$ when electron pairing occurs,
resulting in the destruction of the Dirac double cone $\pm v q$ and the parabolic spectrum $q^2/2m$. Finally, at larger momenta the determinant tends to $-v^2q^4/(2m)$, that is 
the product of the eigenvalues of the Hamiltonian without pairing in the large $q$-limit. 
The determinant and therefore the eigenvalues of the mean-field Hamiltonian depend explicitly on the phase difference of both components of the order parameter. 
It is only in the case of an asymmetric solution $\Delta^{}_1=0$ or $\Delta^{}_2=0$ that the determinant becomes phase independent in the momentum space. 

\begin{figure}[t]
\includegraphics[width=4.5cm]{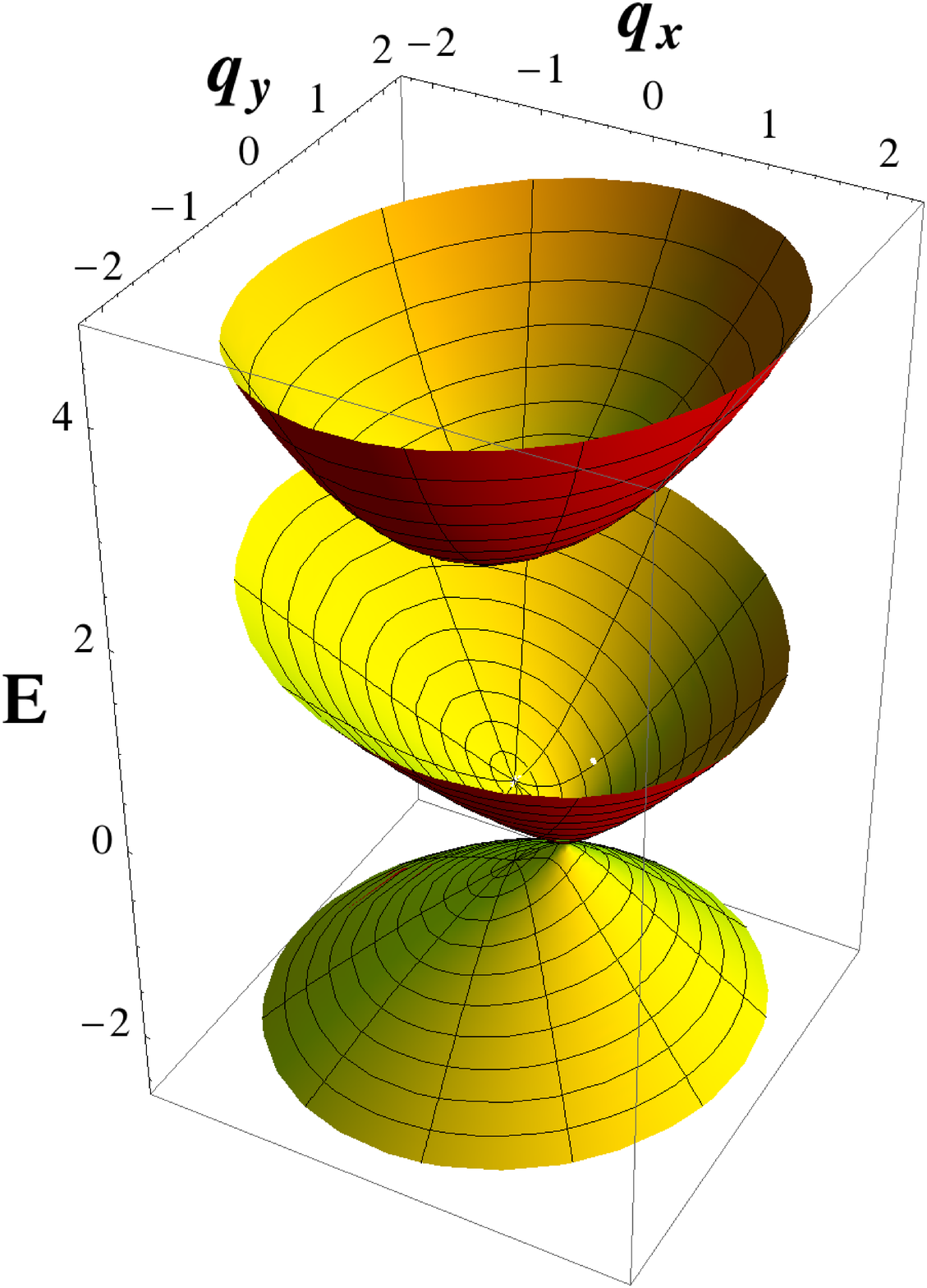}
\includegraphics[width=4.5cm]{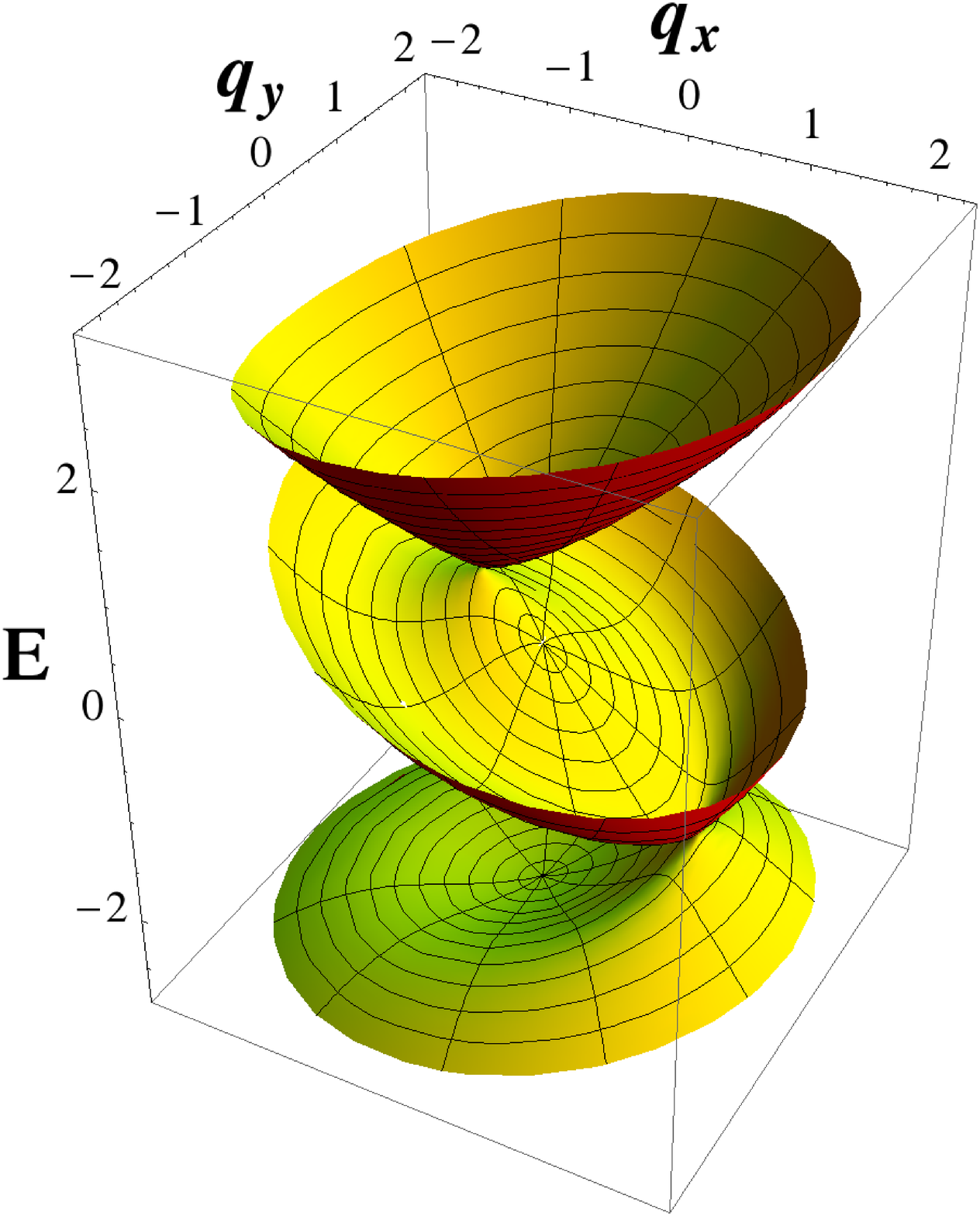}
\includegraphics[width=4.5cm]{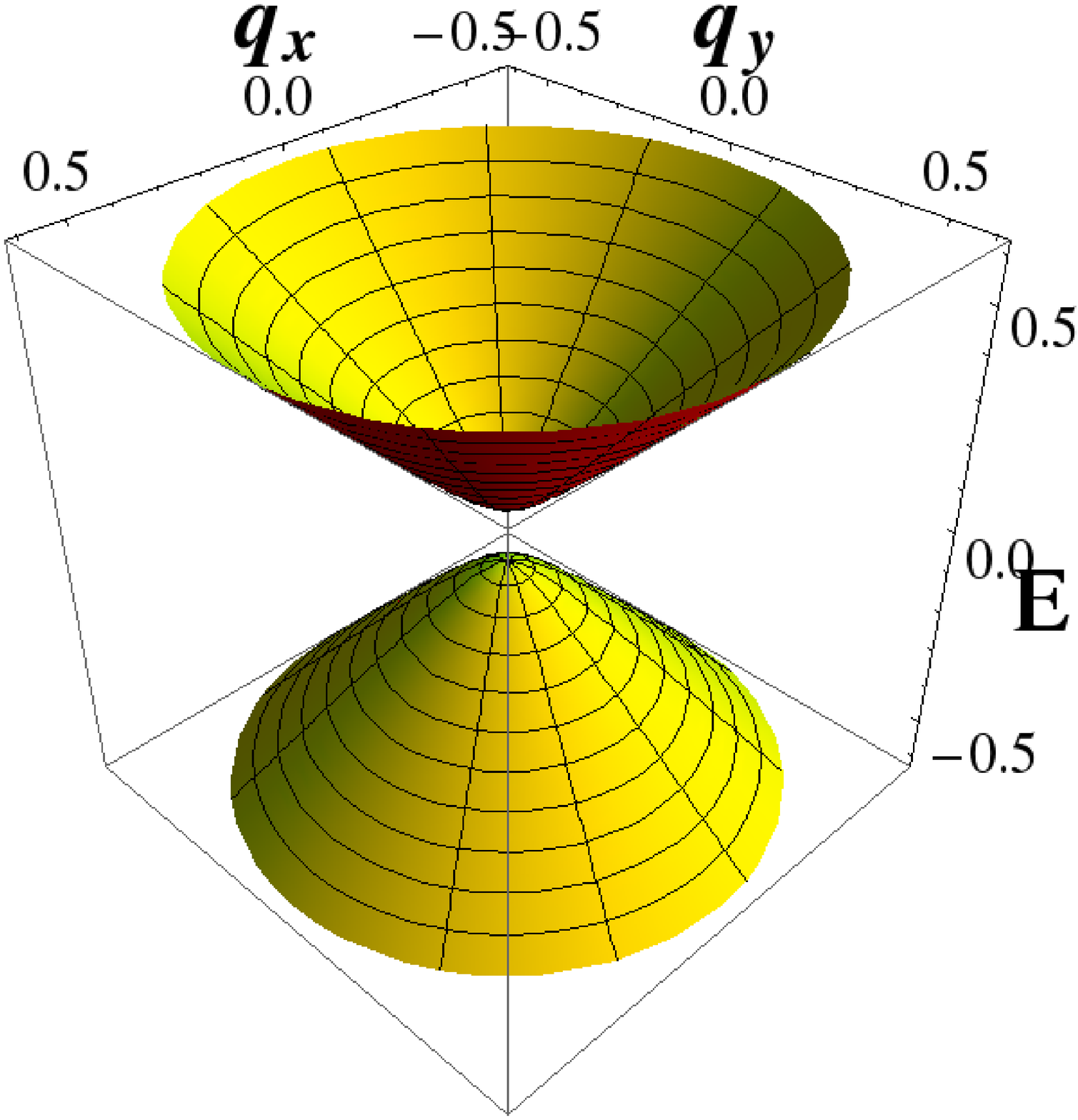}
\caption{
Full band structure of the mean-field Hamiltonian Eq.~(\ref{eq:MFH}) calculated for $mv^2/\Delta=1$.
Left panel: Case without touching points between middle and upper bands according to Eq.~(\ref{eq:cr1}), $\mu/\Delta=-1.5$. 
Middle panel: Case with a touching point between middle and upper bands, $\mu/\Delta=+1.5$. Here, the spectrum is gapless around zero energy.
Right panel: 3d-plot of the quasiparticle spectrum in graphene double layers with paring discussed in Ref.~\cite{SLZ20}
} 
\label{fig:Fig5}
\end{figure}

With variational solutions $\Delta^{}_1=\Delta=\Delta^{}_2$, $\chi^{}_{2} = \chi = \chi^{}_{1}$ and with $q^{}_y=0$ the eigenvalues become:
\begin{equation}
\label{eq:EW}
E^{}_1 = -vq^{}_x, \;\; E^{}_{2} = \frac{1}{2} \left[\xi^{}_q+vq^{}_x + \sqrt{(\xi^{}_q-vq^{}_x)^2 + 8\Delta^2} \right], \;\; 
E^{}_{3} = \frac{1}{2} \left[\xi^{}_q+vq^{}_x - \sqrt{(\xi^{}_q-vq^{}_x)^2 + 8\Delta^2} \right].
\end{equation}
For large {\it positive} $q^{}_x$ (i.e. $\phi=0$) the eigenvalue $E^{}_2$ approaches $\xi^{}_q\sim q^2_x$, thus recovering the dispersion of the conventional particle, 
while the eigenvalue $E^{}_3$ starts at $-(\mu+\sqrt{\mu^2+8\Delta^2})/2$ for small and goes $\sim vq^{}_x\xi^{}_q/\Delta$ for large momenta, 
i.e. it changes from the negative into the positive halfplane. Therefore it will always cross the negative Dirac branch  $E^{}_1$. 
This crossing point represents the deformed original Dirac part of the starting model and is always present in our model irrespective of the particular choice of parameters. 
For small {\it negative} $q^{}_x$ (i.e. $\phi=\pi$) the eigenvalue $E^{}_2$ starts at $(-\mu+\sqrt{\mu^2+8\Delta^2})/2$, goes initially linearly as $\sim -v|q^{}_x|$, 
reaches at some momentum a global minimum and approaches $\xi^{}_q$ for large momenta. 
The eigenvalue $E^{}_3$ approaches $-v|q^{}_x|$ for large momenta, which is the negative branch of the Dirac spectrum, cf Appendix~4. 
Since both $E^{}_1$ and $E^{}_2$ lie at least partially in the positive half-plane, they can cross. 
This is possible if the condition 
\begin{equation}
\label{eq:cr1}
vq^{}_x\geqslant E^{}_2(q^\ast_x,\phi=\pi)
\end{equation}
is fulfilled, where $q^\ast_x$ is the position of the global minimum of $E^{}_2$ in momentum space, which follows from 
\begin{equation}
\left.\frac{\partial}{\partial q^{}_x} E^{}_2(\phi=\pi)\right|_{q^{}_x=q^\ast_x} = 0. 
\end{equation}
In the case of strict equality in Eq.~(\ref{eq:cr1}) there is only one single touching point between both bands, while for "$>$" we have two crossing points, cf. Appendix~4.2. 
The crossing at higher energies has the shape of a single point only along the projection $q^{}_y=0$. In general it is an extended curve while 
the crossing at lower energies remains a single point.

\section{The nematicity of the full spectrum  of the Hamiltonian~(\ref{eq:MFH})} 

\begin{figure}[t]
\includegraphics[width=4.75cm]{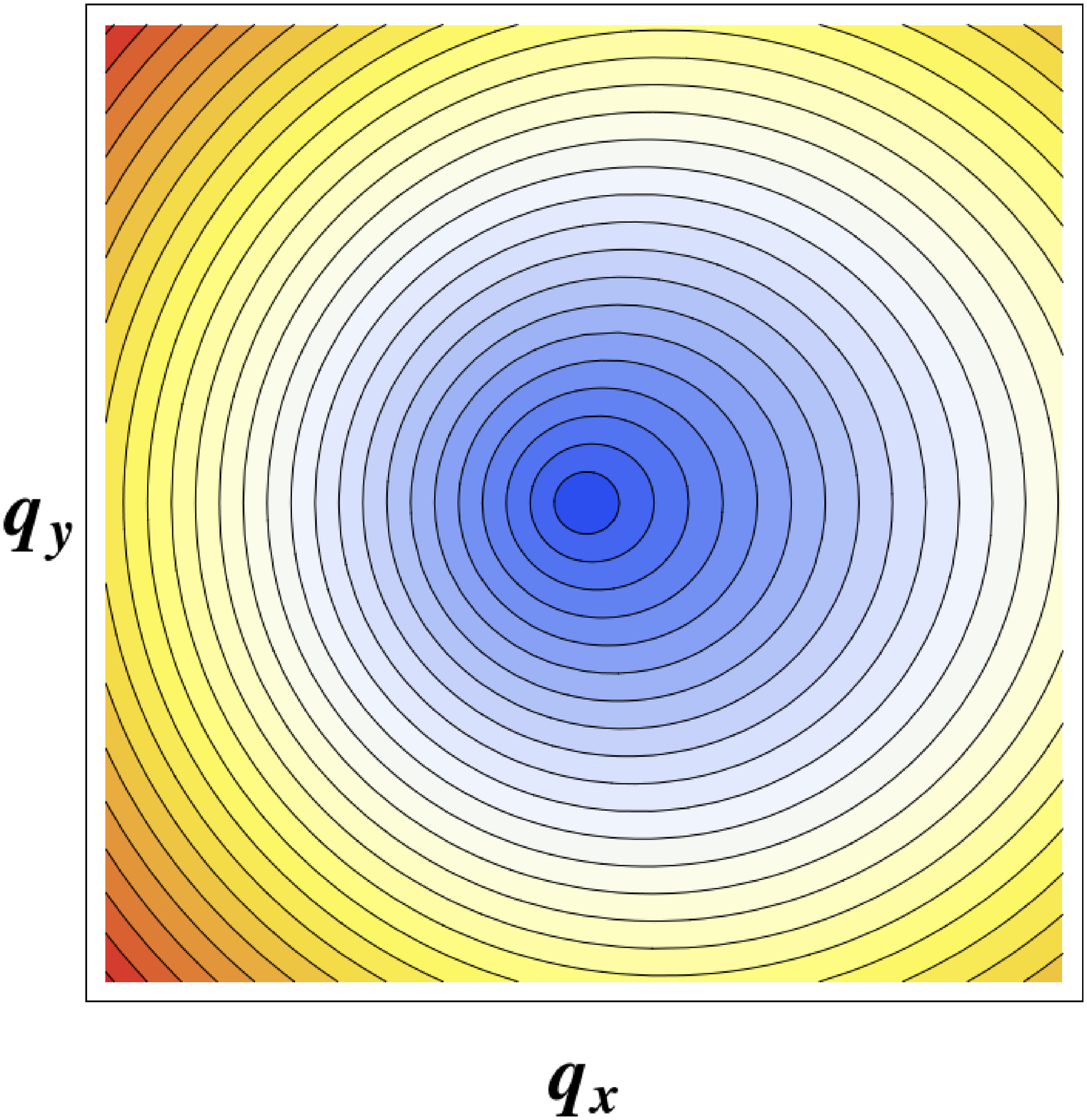}
\hspace{4mm}
\includegraphics[width=4.75cm]{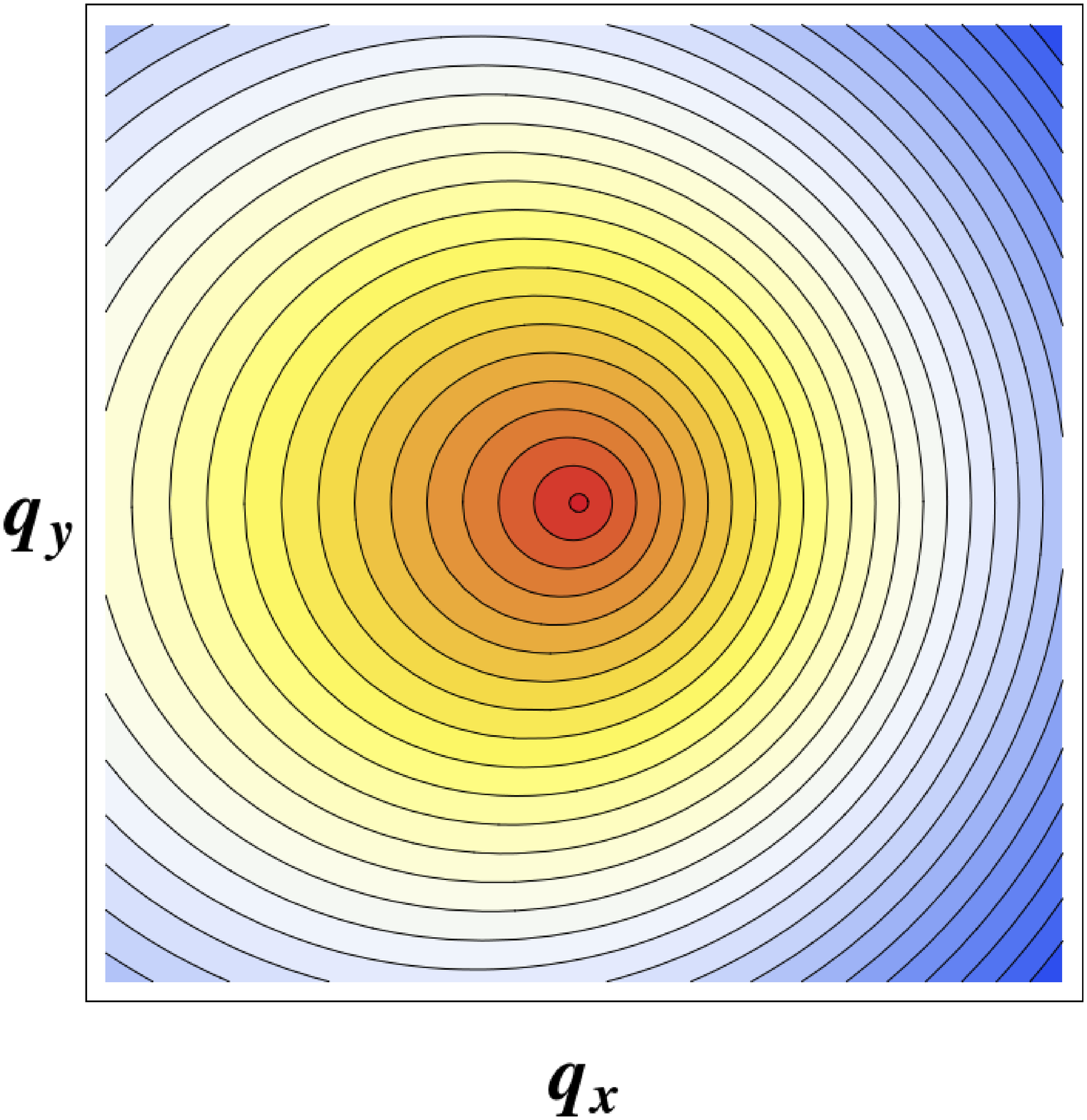}

\vspace{2mm}
\includegraphics[width=4.75cm]{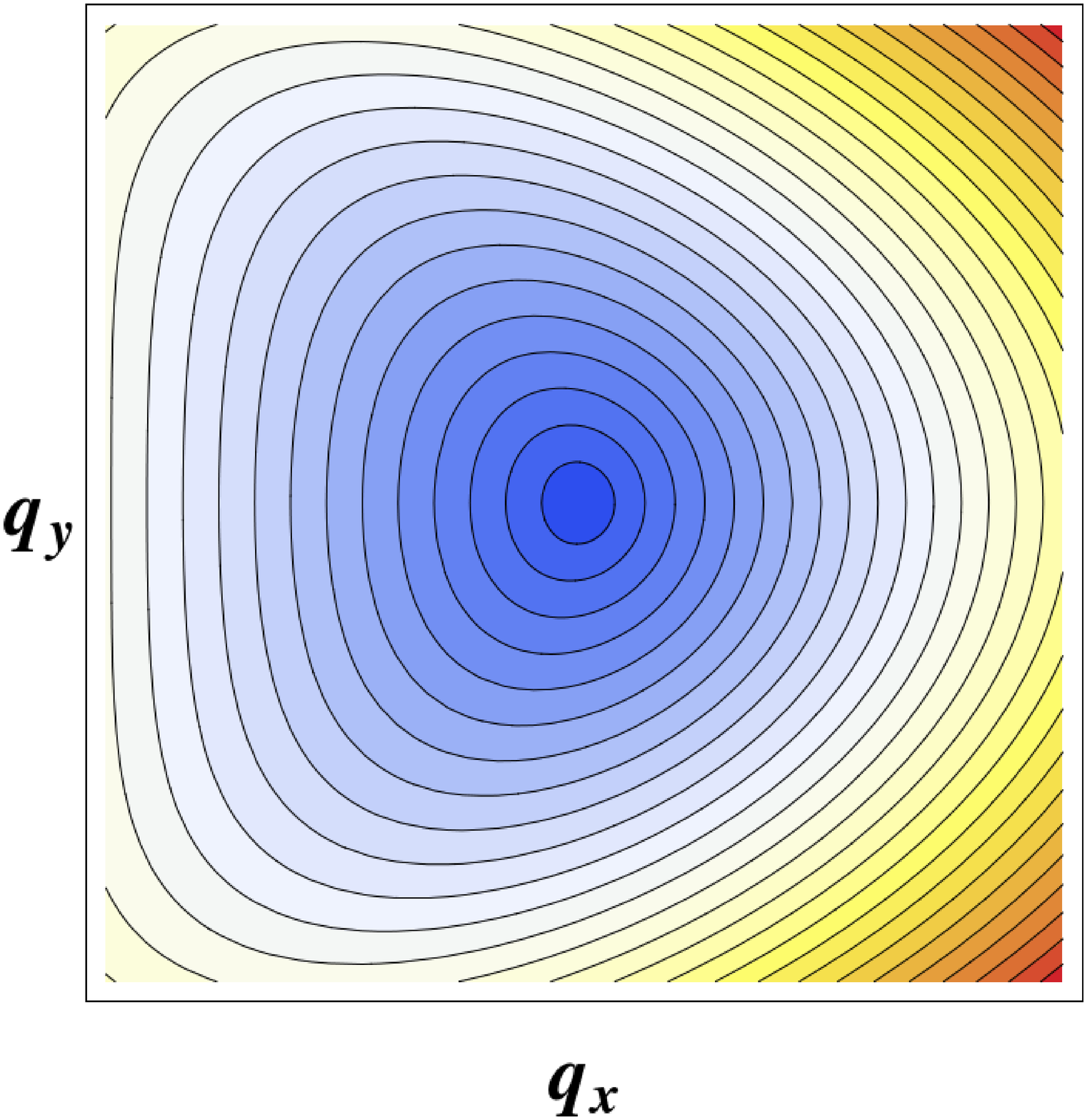}
\hspace{4mm}
\includegraphics[width=4.75cm]{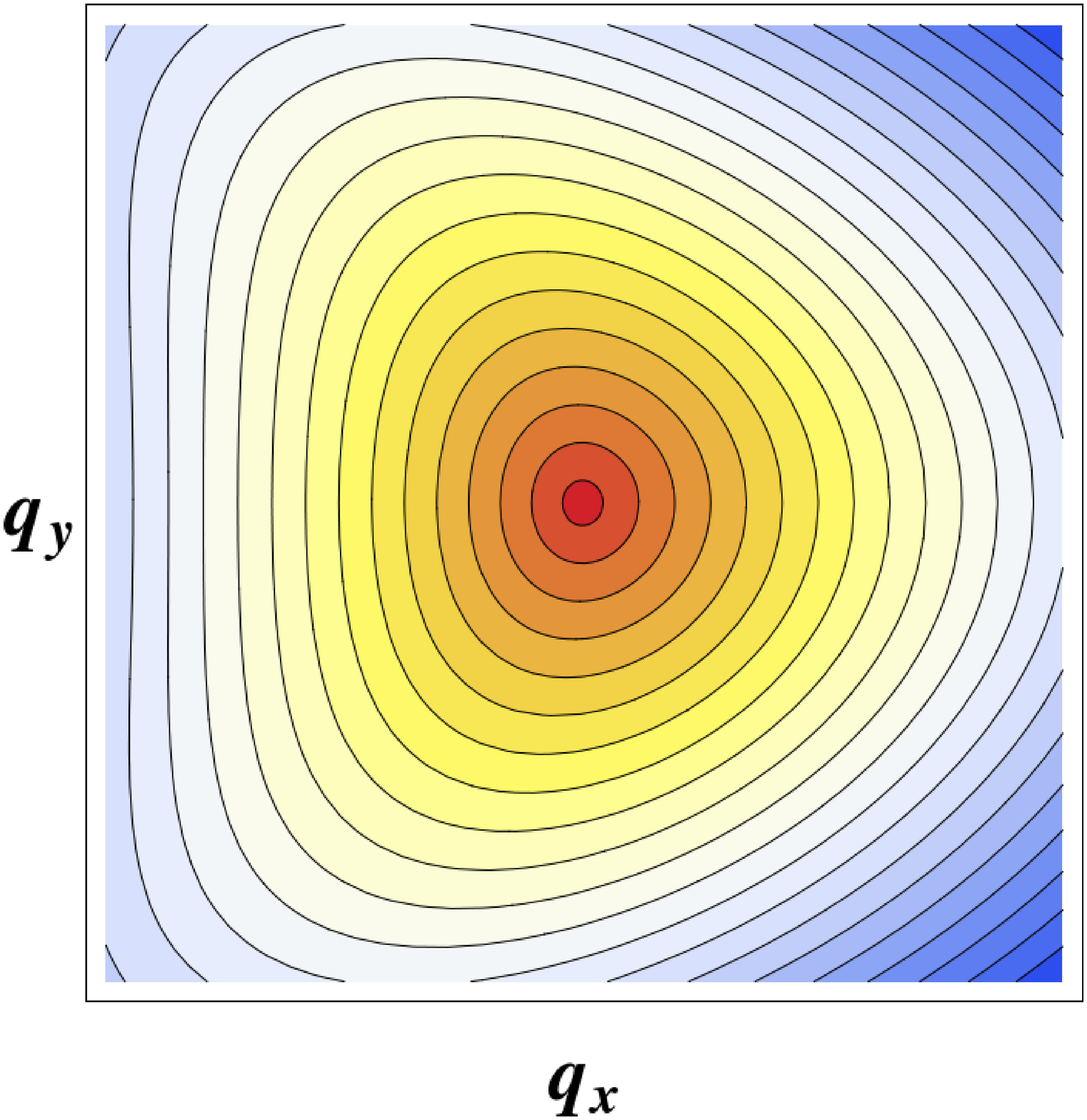}
\caption{
The spontaneous nematicity of the quasiparticle spectrum in the closest vicinity of both nodal points shown in the middle panel of Fig.~\ref{fig:Fig5}.
Upper (left) and lower (right) half-cone of the upper (upper row) and lower (bottom row) cone are shown. 
Due to anisotropy they are not concentric circles but rather a kind of deformed surfaces jammed in $q^{}_x$-direction.
Darker colors correspond to the lower energies in each graphic. 
}
\label{fig:Fig6}
\end{figure}

The eigenvalues of the Hamiltonian~(\ref{eq:MFH}) for zero Dirac mass are found by solving the cubic equation and represent the Cardano formulae, cf. Appendix~4.1.
In Fig.~\ref{fig:Fig5} we show two cases: The case without a crossing point between the middle and upper band and the case with such a point, as it is discussed in the previous paragraph. 
In both discussed cases the system parameters are intentionally chosen large to emphasize the characteristic features of the spectral bands and DOS, cf.  Fig.~\ref{fig:Fig5}. 
In the first case we have effectively a system of original Dirac and conventional particles with only slight spectral deformation and a shift in momentum space due to the pairing. 
The case with touching between the both upper bands is more involved. 
In this case the spectrum is characterized by a strong spatial anisotropy at every point in the momentum space. 
Such spectral anisotropy which occurs due to interaction and does not break the translational symmetry of the lattice is sometimes called the spontaneous nematicity~\cite{Fradkin2010,Fernandes2014,Fernandes2019,cao20}.
In particular, around zero energy the spectrum is linear in $q^{}_x$-direction with the slope given by the Fermi velocity of the original Dirac particle $v^{}_x=v$ , 
while in $q^{}_y$-direction a parabolic spectrum with negative momentum dependent Fermi velocity $v^{}_y=-\mu q^{}_y/{2m\Delta}$.
The touching points are complex structures with both bands penetrating each other. 
In the lowest band one recognizes a smooth transition between regions with positive and negative curvatures, which results in formation of a local energy minimum.
In the close vicinity of the crossing points the spectrum of both involved bands has the shape of deformed Dirac cones, Fig.~\ref{fig:Fig6}.
For larger  momenta the energy bands unbind, the phase dependence of eigenvalues disappears, and the spectra of decoupled Dirac and conventional 
particles are recovered. The spectra in Fig.~\ref{fig:Fig5} are plotted for the fixed phase difference of order parameters $\chi^{}_2-\chi^{}_1=0$. 
The spectrum rotates if $\chi^{}_2-\chi^{}_1$ changes as pointed out in the discussions of the anisotropy around Eq.~(\ref{eq:detGmf}).

\section{The density of states of the Hamiltonian~(\ref{eq:MFH})} 

\begin{figure}[t]
\includegraphics[width=4.cm]{Fig5a.eps}
\hspace{5mm}
\includegraphics[width=7.75cm]{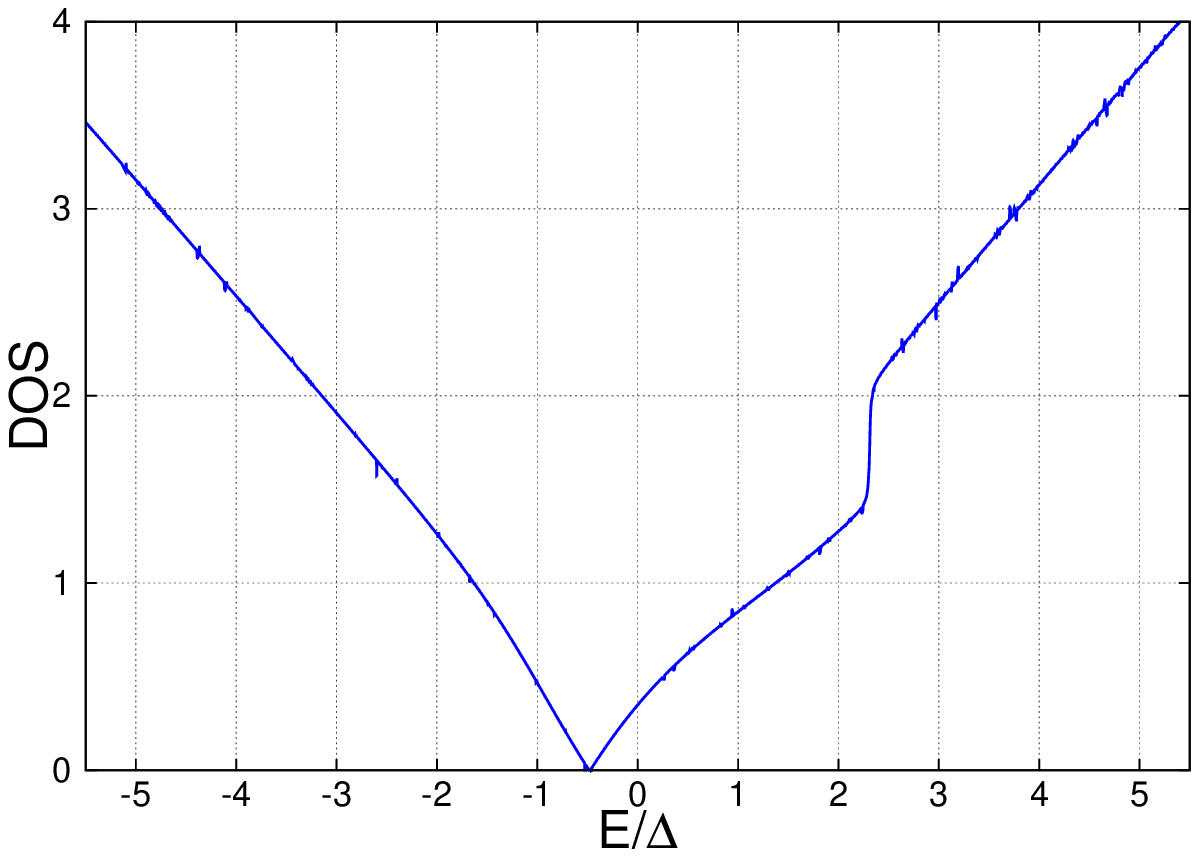}

\includegraphics[width=4.5cm]{Fig5b.eps}
\hspace{5mm}
\includegraphics[width=7.75cm]{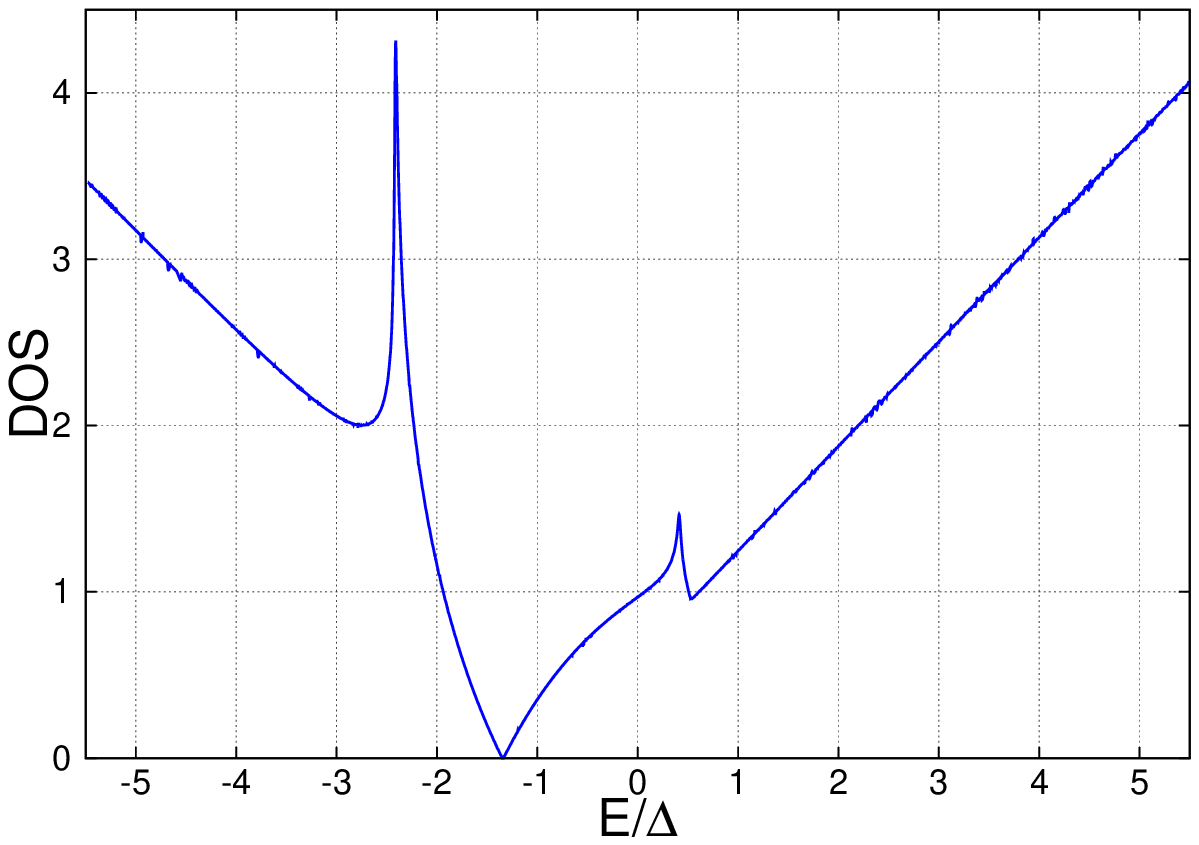}
\caption{
The DOS in arbitrary units of the mean-field Hamiltonian Eq.~(\ref{eq:MFH}) vs. the corresponding spectrum for the same set of parameters . 
Upper figure: Regime without touching points between middle and upper bands. 
Lower figure: Regime with a touching point between middle and upper bands. 
The sharp peak between $-3\Delta$ and $-2\Delta$ is due to saddle point which occur in the lowest band. 
The structure at around $0.5\Delta$ is due to the touching between two higher bands. 
}
\label{fig:Fig7}
\end{figure}

We evaluate the density of states (DOS) of the mean-field Hamiltonian from the usual functional
\begin{equation}
{\rm DOS}(E)  = {\rm Tr}\delta(E-H^{}_{\rm MF}),
\end{equation}
where $E$ is the energy and the momentum summation if performed over the rectangular Brillouin zone. 
The DOS for both spectra depicted in Fig.~\ref{fig:Fig5} are shown in Fig.~\ref{fig:Fig7}.
For the case without touching between two upper bands the DOS shows nicely the low lying slightly deformed Dirac part up to the energies of roughly 2.5$\Delta$ where the uppermost
parabolic band is reached. This is recognizable by the sharp jump in DOS, since the DOS of parabolic spectrum in 2d is the Heaviside step function. 
In the case with the touching between both upper bands the DOS structure is more complex. 
We still recognize the Dirac particle at energies below $-3\Delta$ and above $+\Delta$, as well as the remnants of the strongly deformed original 
Dirac cone at roughly $-1.5\Delta$. However, here appears a sharp peak corresponding to a highly populated state between $-2\Delta$ and $-3\Delta$, 
which is absent in the (almost) decoupled case. This state is due to the strong deformation of the lowest band visible in Fig.~\ref{fig:Fig5}. 
The peak in DOS is therefore a van-Hove singularity due to the saddle point which forms in the lowest band due to increasing pairing order parameter, cf. Appendix 4. 
The presence of the touching between the middle and upper bands manifests itself in the DOS by the structure visible around $0.5\Delta$. 
The initial increment in the DOS is here due to the parabolic band which penetrates the positive Dirac band as discussed
around Eq.~(\ref{eq:EW}). The DOS increases up to the crossover energy at which the scaling of the upper band changes from parabolic to linear. 
From this energy the DOS decreases up to the position of the Dirac-like band touching point.
For higher energies, the DOS is again that of the asymptotic Dirac spectrum. Second crossing between the two bands at higher energies does not show up in the DOS, 
i.e. nothing significant occurs here and both bands simply go through each other retaining the same scaling.

\section{Discussions} 

In this paper we study a model with a single spinless Dirac fermion in one layer and a single spin projection of conventional electrons in the other. 
The mean-field Hamiltonian in Eq.~(\ref{eq:MFH}) is a 3$\times$3 matrix, which consequently has three eigenvalues with the strong spatial anisotropy. 
The spectrum of the Hamiltonian is gapless and linear at energies around zero. 
This is very different from the case of conventional s-wave superconductivity in graphene double layers, which we investigated in  Ref.~\cite{SLZ20}.
In that case the spectrum has the usual shape of two paraboloids separated by the spectral gap around zero energy as shown on the right side of Fig.~\ref{fig:Fig5}.
Another consequence of the spectral anisotropy is the formation of the band crossing points due to deformation of original Dirac and parabolic spectra. 
In the vicinity of band crossing points, the spectrum has the form of deformed Dirac cones.
The existence of the band crossing should be detectable by standard experimental spectroscopic techniques. 
The deformation of the bands leads to appearance of van Hove singularities in the density of states.

The formation of the pairing state is influenced by an interplay of different model's parameters such as the Dirac mass and the Fermi velocities of both fermion species. 
The critical interaction strength of the pairing transition is a function of the Fermi velocities of both electron species. 
The effective potential which is associated with the pairing transition turns out to be a complicated function of system's parameters. 
The effective potential exhibits pronounced local minima. We identify several factors which determine the position of these minima.

\section{Conclusions and outlook} 
In the system consisting of two layers with Dirac and conventional electrons respectively, there is an interlayer electron-electron pairing transition due to strong interlayer repulsion. 
The main concern of this paper are the studies of the spectrum of quasiparticles and details of the pairing transition.
Several problems are left for the future, e.g. the intra- and interlayer transport in such layered systems, for which the fluctuations beyond the Gaussian order are necessary, 
or physics of time-reversal symmetry breaking by e.g. external magnetic fields or sharp sample boundaries, cf. Ref.~\cite{ZSL21}. 

\section{Acknowledgments} 
This research was supported by a grant of the Julian Schwinger Foundation for Physics Research.
Yu.E.L. was supported by the basic research program of the national research university Higher School of Economics
A.S. was partially supported by the German Research Foundation through Transregio TRR80.

\appendix
\section{Appendix}

\subsection{1. Derivation of the mean field Hamiltonian~(\ref{eq:MFH}) from the microscopic model with contact density-density interaction}
\label{app:proof}

Here we consider a version of the microscopic Hamiltonian (\ref{eq:MFH}) with the Coulomb interaction modeled by the simplest contact density-density interaction. 
Formally we model both species as if they were confined to the same layer. The justification of this approximation is given in Ref.~\cite{SLZ20} and references therein. 
Arguably, the local approximation for the interlayer interaction gives quantitatively the crudest estimation, but does not change the qualitative picture~\cite{Stoof2011,Berman2019,SLZ20,MacDonald2020,Fogler2014}. The model Hamiltonian defined in this way reads
\begin{equation}
\label{eq:TrialHam}
{\cal H} = \psi^\dag\cdot[- iv\nabla\cdot\sigma + \Delta^{}_{\rm D}\sigma^{}_3]\psi + \varphi^\dag\cdot[ -\frac{\nabla^2}{2 m} - \mu]\varphi + \frac{g}{2}(\psi^\dag\psi)(\varphi^\dag\varphi),
\end{equation}
where $g$ is the interaction strength and the rest of the parameter set is already defined in the main part. The kinetic part represents a 3$\times$3 quadratic form
\begin{equation}
{\cal H}^{}_0 = 
\left(
\begin{array}{c}
\psi^\dag_1 \\
\psi^\dag_2 \\
\varphi^\dag
\end{array}
\right)\cdot
\left(
\begin{array}{ccc}
  \Delta^{}_{\rm D}  &  -iv\nabla^{}_1 - v\nabla^{}_2     &      0         \\  
 -iv\nabla^{}_1 + v\nabla^{}_2       &   -\Delta^{}_{\rm D} &      0         \\
                 0                   &                0                   & -\frac{\nabla^2}{2 m} - \mu          
\end{array}
\right)
\left(
\begin{array}{c}
\psi^{}_1 \\
\psi^{}_2 \\
\varphi^{}
\end{array}
\right),
\end{equation}
which gives a hint in which position in the matrix the order parameter can appear. To see if this guess is compatible with the form of the interaction at hand we rewrite it in the following way
\begin{equation}
\label{eq:rep}
(\psi^\dag\psi)(\varphi^\dag\varphi) \sim - \frac{1}{2}\sum^{3}_{j=0}\left(\psi^\dag\sigma^{}_j\vec v^{}_\u \varphi\right)\left(\varphi^\dag\vec v^t_\u\sigma^{}_j\psi\right),
\end{equation}
where $\vec v^t_\u = (1,0)$ and $\sigma^{}_0$ the 2d unity matrix. The operators $\psi$ and $\varphi$ anticommute and single operator bilineals are neglected since they do not contribute to the pairing and are generally the artifact of this operator ordering. 
To prove our guess, the bilineal $\varphi^\dag\varphi$ can be pulled out the sum over $j$ in Eq.~(\ref{eq:rep}) with the change of the sign due to fermionic statistics.
Straightforward calculations yield
\begin{equation}
\left(\psi^\dag\sigma^{}_0\vec v^{}_\u \right) = \left(\psi^\dag\sigma^{}_3\vec v^{}_\u \right) = \psi^\dag_1, 
\;\; \left(\psi^\dag\sigma^{}_1\vec v^{}_\u \right) = \psi^\dag_2;\;\; \left(\psi^\dag\sigma^{}_2\vec v^{}_\u \right) = i\psi^\dag_2,
\end{equation}
and 
\begin{equation}
\left(\vec v^t_\u\sigma^{}_0\psi\right) = \left(\vec v^t_\u\sigma^{}_3\psi\right) = \psi^{}_1,
\;\;\left(\vec v^t_\u\sigma^{}_1\psi\right) = \psi^{}_2,\;\;\left(\vec v^t_\u\sigma^{}_2\psi\right) = -i\psi^{}_2.
\end{equation}
Combining them as in Eq.~(\ref{eq:rep}) yields the initial interaction term. So, we don't loose any parts by this reordering.
We therefore can introduce a full set of order parameters allowed by the construction of the interaction
\begin{eqnarray}
\Delta^{}_\mu   = -\frac{g}{4}\langle \psi^\dag\sigma^{}_\mu\vec v^{}_\u\varphi\rangle = \frac{g}{4}{\rm tr}~\sigma^{}_\mu \langle\varphi \vec v^{}_\u\otimes\psi^\dag\rangle, \\
\Delta^\ast_\mu = -\frac{g}{4}\langle \varphi^\dag\vec v^t_\u\sigma^{}_\mu\psi\rangle  = \frac{g}{4}{\rm tr}~\sigma^{}_\mu \langle \psi\otimes\vec v^t_\u\varphi^\dag\rangle.
\end{eqnarray}
In particular, each of the order parameters reads
\begin{eqnarray}
\Delta^{}_0 = \Delta^{}_3 = \frac{g}{4} \langle\varphi\psi^\dag_1\rangle,\,\,\, \Delta^{}_1 = \frac{g}{4} \langle\varphi\psi^\dag_2\rangle,\,\, \Delta^{}_2 = -i\frac{g}{4} \langle\varphi\psi^\dag_2\rangle .
\end{eqnarray}
Obviously, these correlators can only be finite in the paired state.

\subsection{2. Evaluation of the critical interaction strength from Eq.~(\ref{eq:CritCond})} 
\label{app:gc}

At zero temperature, the integration of the frequency in Eq.~(\ref{eq:CritCond}) can be performed most comfortably by the residue theorem. 
When exploiting the logic of the modulus operator we get
\begin{eqnarray}
\frac{1}{2g^{}_c} &=& \int^\Lambda\frac{d^2q}{(2\pi)^2} \intop^\infty_{-\infty}\frac{dq^{}_0}{2\pi}~\frac{q^2_0}{(q^2_0+v^2q^2)(q^2_0+\xi^2_q)} = 
\frac{1}{8\pi}\int^\Lambda_0 dq~\frac{q}{vq+\left|\frac{q^2-q^2_F}{2m}\right|} \\
&=& \frac{1}{8\pi}\int^\Lambda_0 dq\left[\Theta(q-q^{}_F)\frac{q}{vq+\frac{q^2}{2m}-\frac{q^2_F}{2m}} + \Theta(q^{}_F-q)\frac{q}{vq - \frac{q^2}{2m}+\frac{q^2_F}{2m}}\right] \\
&=& \frac{1}{8\pi}\int^\Lambda_{q^{}_F} dq ~\frac{q}{vq+\frac{q^2}{2m}-\frac{q^2_F}{2m}} 
+ \frac{1}{8\pi}\int^{q^{}_F}_0dq ~\frac{q}{vq - \frac{q^2}{2m}+\frac{q^2_F}{2m}} \\
&=& \frac{m}{4\pi}\int^\Lambda_{q^{}_F} dq ~\frac{q}{q^2+2mvq-q^2_F} + \frac{m}{4\pi}\int^{q^{}_F}_0dq ~\frac{q}{q^2_F+2mvq-q^2},
\end{eqnarray}
or upon reordering and introducing dimensionless quantities $x=q/q^{}_F$, $y=mv/q^{}_F$, and $\lambda=\Lambda/q^{}_F$ 
\begin{eqnarray}
\frac{2\pi}{mg^{}_c} &=& \intop^\lambda_{1} dx ~\frac{x}{x^2+2yx-1} - \intop^{1}_0 dx~\frac{x}{1+2yx-x^2} \\
&=& \int^\lambda_1dx~\frac{x}{(x+a^{}_-)(x+a^{}_+)} - \int^1_0dx~\frac{x}{(x-a^{}_-)(x-a^{}_+)},
\end{eqnarray}
where $a^{}_{\pm} = y \pm \sqrt{1+y^2}$. Utilizing the partial fraction decomposition and performing the integrals we get to the expression Eq.(\ref{eq:critg}).

\subsection{3. Evaluation of the effective potential}

\subsection{3.1 Zeroth order term in gradient expansion}

\label{app:ZerothOrd}

\begin{figure}[t]
\includegraphics[width=7.5cm]{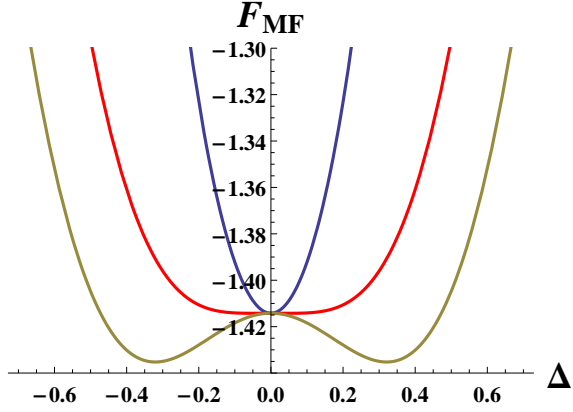}
\caption{(Color online)
Three profiles of the effective potential as function of $\Delta$ for different effective interaction. The critical interaction strength is $\gamma^{}_c=2\sqrt{2}$. 
}
\label{fig:Fig1S}
\end{figure}

The effective potential to zeroth order in gradient expansion reads
\begin{equation}
{\cal F}^{(0)}_{MF} \approx \frac{1}{2g}\left(\Delta^2_1+\Delta^2_2\right) - \frac{\Lambda^2}{4\pi}\int\frac{dq^{}_0}{2\pi}~ 
\log\left[ 
-iq^{}_0 (q^2_0+\Delta^2_1+\Delta^2_2) + \mu q^2_0 
\right],
\end{equation}
where we trivially integrated the momentum. We express the order parameter in units of the chemical potential 
\begin{equation}
\Delta^2_1+\Delta^2_2 = \mu^2\Delta^2 ,
\end{equation}
which leads us to 
\begin{equation}
\label{eq:MF0}
{\cal F}^{(0)}_{MF} \approx \frac{\mu^2\Delta^2}{2g} - \frac{\Lambda^2}{4\pi}\intop^{\infty}_{-\infty}\frac{dq^{}_0}{2\pi}~ 
\log\left[ -iq^{}_0 (q^2_0+\mu^2\Delta^2) + \mu q^2_0 \right].
\end{equation}
In order to perform the momentum integral we write the logarithm as 
\begin{eqnarray}
\intop^{\infty}_{-\infty}\frac{dq^{}_0}{2\pi}\log\left[ -iq^{}_0 (q^2_0+\mu^2\Delta^2) + \mu q^2_0 \right] &=&
\intop^{\infty}_{-\infty}\frac{dq^{}_0}{2\pi}\int d\Delta\frac{\partial}{\partial\Delta}\log\left[-iq^{}_0 (q^2_0+\mu^2\Delta^2) + \mu q^2_0 \right] \\
=\int d\Delta^2\intop^{\infty}_{-\infty}\frac{dq^{}_0}{2\pi}\frac{-iq^{}_0\mu^2}{-iq^{}_0 (q^2_0+\mu^2\Delta^2) + \mu q^2_0 } 
&=& \int d\Delta^2\intop^{\infty}_{0}\frac{dq^{}_0}{2\pi}\frac{2\mu^2 (q^2_0+\mu^2\Delta^2)}{(q^2_0+\mu^2\Delta^2)^2 + \mu^2 q^2_0}.
\end{eqnarray}
Performing the partial fraction decomposition we further get using
\begin{equation}
\label{eq:Roots}
E^{}_\pm = \frac{\mu^2}{2} \left(1+2\Delta^2\pm\sqrt{1+4\Delta^2}\right)
\end{equation}
the following expression
\begin{eqnarray}
&&2\mu^2\int d\Delta^2\intop^{\infty}_{0}\frac{dq^{}_0}{2\pi}\frac{2\mu^2 (q^2_0+\mu^2\Delta^2)}{(q^2_0+\mu^2\Delta^2)^2 + \mu^2 q^2_0}  \\
&=& 
2\mu^2\int d\Delta^2\intop^{\infty}_{0}\frac{dq^{}_0}{2\pi}
\left[ 
\frac{E^{}_+-\mu^2\Delta^2}{E^{}_+-E^{}_-}\frac{1}{q^2_0+E^{}_+} - \frac{E^{}_--\mu^2\Delta^2}{E^{}_+-E^{}_-}\frac{1}{q^2_0+E^{}_-} 
\right]\\
&=& 
2\mu^2\int d\Delta^2\left[\frac{1}{\sqrt{E^{}_+}} \frac{E^{}_+-\mu^2\Delta^2}{E^{}_+-E^{}_-} - \frac{1}{\sqrt{E^{}_-}}\frac{E^{}_--\mu^2\Delta^2}{E^{}_+-E^{}_-} \right] \\
&=&
\sqrt{2}\mu \int d\Delta^2
\left[
\frac{\sqrt{1+4\Delta^2}+1}{\sqrt{1+4\Delta^2}\sqrt{1+2\Delta^2+\sqrt{1+4\Delta^2}}} + \frac{\sqrt{1+4\Delta^2}-1}{\sqrt{1+4\Delta^2}\sqrt{1+2\Delta^2-\sqrt{1+4\Delta^2}}}
\right] \\
&=& \sqrt{2}\mu\left(\epsilon^{\frac{1}{2}}_+ + \epsilon^{\frac{1}{2}}_-\right),
\end{eqnarray}
where 
\begin{equation}
\epsilon^{}_\pm = 1+2\Delta^2 \pm \sqrt{1+4\Delta^2}.
\end{equation}
Therefore the effective potential Eq.~(\ref{eq:MF0}) becomes 
\begin{equation}
\label{eq:MF01}
{\cal F}^{(0)}_{MF} \approx \frac{\mu\Lambda^2}{2\sqrt{2}\pi}
\left[ 
\frac{\Delta^2}{\gamma} - \left(\epsilon^{\frac{1}{2}}_+ + \epsilon^{\frac{1}{2}}_- \right)
\right]
\end{equation}
with the effective dimensionless interaction strength
\begin{equation}
\gamma = \frac{g\Lambda^2}{\sqrt{2}\pi\mu} 
\end{equation}
For $\gamma>\gamma^{}_c=2\sqrt{2}$ the effective potential develops the local minima corresponding to the paired state, cf. left panel of Fig.~\ref{fig:Fig1S}. 
The variation of this expression with respect to the gap parameter yields a condition for the critical gap parameter, which is plotted in the 
right panel of Fig~\ref{fig:Fig1S}. As predicted for $\gamma\geqslant2\sqrt{2}$ there are non-trivial zero solutions for of the variational equation. 

\subsection{3.2 Effect of the small Dirac mass on effective energy functional}
\label{app:SOmass}

Now we include the Dirac mass and study how it may change the shape of the effective energy to the considered approximation, 
whereas it suffices to only consider the effect of the mass in leading order of perturbative expansion in $\Delta^{}_{\rm D}$. 
The free energy functional changes as
\begin{eqnarray}
\nn
{\cal F}^{(1)}_{MF} &\approx& \frac{1}{2g}\left(\Delta^2_1+\Delta^2_2\right) - \frac{\Lambda^2}{4\pi}\int\frac{dq^{}_0}{2\pi}~ 
\log\left[ 
-iq^{}_0 (q^2_0+\Delta^2_1+\Delta^2_2) + \mu q^2_0 +  \Delta^{}_{\rm D}(\Delta^2_1-\Delta^2_2)
\right],
\end{eqnarray}
and the next power of $\Delta^{}_{\rm D}$ under logarithm being quadratic. We recognize that the small Dirac mass breaks the sublattice-symmetry between the components of the pairing gap. 
The leading order effect is captured by expansion of the logarithm part
\begin{eqnarray}
\nn
{\cal F}^{(1)}_{MF} &\approx& \frac{1}{2g}\left(\Delta^2_1+\Delta^2_2\right) - \frac{\Lambda^2}{4\pi}\int\frac{dq^{}_0}{2\pi}~ 
\log\left[ 
\mu q^2_0 - iq^{}_0 (q^2_0+\Delta^2_1+\Delta^2_2)
\right] \\
&& 
- \frac{\Lambda^2}{4\pi}\int\frac{dq^{}_0}{2\pi}~  
\frac{\Delta^{}_{\rm D}(\Delta^2_1-\Delta^2_2)}{\mu q^2_0-iq^{}_0 (q^2_0+\Delta^2_1+\Delta^2_2)},
\end{eqnarray}
We evaluate the integral in the correction term:
\begin{eqnarray}
\intop^{\infty}_{-\infty}\frac{dq^{}_0}{2\pi}~\frac{1}{\mu q^2_0-iq^{}_0 (q^2_0+\Delta^2_1+\Delta^2_2)} &=& 
\intop^{\infty}_{-\infty}\frac{dq^{}_0}{2\pi}~\frac{\mu}{\mu^2 q^2_0 + (q^2_0+\Delta^2_1+\Delta^2_2)^2} \\
= \intop^{\infty}_{-\infty}\frac{dq^{}_0}{2\pi}~\frac{\mu}{[q^2_0+E^{}_+][q^2_0+E^{}_-]} &=& 
\frac{\sqrt{2}}{\mu^2}\frac{1}{\sqrt{1+4\bar\Delta^2_1+4\bar\Delta^2_2}}
\frac{\epsilon^{\frac{1}{2}}_+-\epsilon^{\frac{1}{2}}_-}{\epsilon^{\frac{1}{2}}_-\epsilon^{\frac{1}{2}}_+}.
\end{eqnarray}
where $E^{}_{\pm}$ are taken over from Eq.~(\ref{eq:Roots}) and all energy quantities are rescaled in units of chemical potential $\mu$. 
Taking all terms together we get the free-energy functional Eq.~(\ref{eq:MF0}) becomes Eq.~(\ref{eq:MF02}):
\begin{equation}
\label{eq:MF02}
{\cal F}^{(1)}_{MF} \approx \frac{\mu\Lambda^2}{2\sqrt{2}\pi}
\left[ 
\frac{\bar\Delta^2_1+\bar\Delta^2_2}{\gamma} - 
\left(\epsilon^{\frac{1}{2}}_+ + \epsilon^{\frac{1}{2}}_- + \frac{\bar\Delta^{}_{\rm D}(\bar\Delta^2_1-\bar\Delta^2_2)}
{\sqrt{1+4\bar\Delta^2_1+4\bar\Delta^2_2}}
\frac{\epsilon^{\frac{1}{2}}_+-\epsilon^{\frac{1}{2}}_-}{\epsilon^{\frac{1}{2}}_-\epsilon^{\frac{1}{2}}_+}
\right)
\right]
\end{equation}
Since the sign of the Dirac mass is not fixed as such, the Dirac mass breaks explicitly the sublattice symmetry of the order parameter.

\subsection{3.3 Approximate treatment of the effective potential for small momenta}

For very small momenta $(q^2\ll q^2_F)$ and in absence of the Dirac mass we can approximate the mean-field potential Eq.~(\ref{eq:fef}) as 
\begin{equation}
\label{eq:log}
{\cal F}^{(2)}_{MF} \approx \frac{\Delta^2_1+\Delta^2_2}{2g} - \int\frac{d^3Q}{(2\pi)^3}~ 
\log\left[ 
-iq^{}_0 (q^2_0+\Delta^2_1+\Delta^2_2) + \mu q^2_0  + 2 vq\Delta^{}_1\Delta^{}_2 \cos\left(\chi^{}_{2} - \chi^{}_{1} + \phi\right)
\right].
\end{equation}
The suggested smallness of the momentum enables us to use it as an expansion parameter and consequently to perform the angular integration with respect to the angle $\phi$. 
This makes all odd-power contributions vanish. The resulting expression is rotationally invariant and isotropic with respect to both the momentum and the order-parameter, 
i.e. the phase of the order parameter disappear from the final result as consequence of the 'gauge invariance'. Using the notation   
$\displaystyle  X = -iq^{}_0 (q^2_0+\Delta^2_1+\Delta^2_2) + \mu q^2_0 $ and  $\displaystyle  Y = v\Delta^{}_1\Delta^{}_2$ we have 
\begin{eqnarray}
\label{eq:line2}
\intop^{2\pi}_0\frac{d\phi}{2\pi}~\log[X+2Y\cos\left(\chi^{}_{2} - \chi^{}_{1} + \phi\right)]  
&=& \log[X] - \sum^\infty_{n=1}q^{2n}\frac{2^{n-1}}{n}\frac{(2n-1)!!}{n!} \left(\frac{Y}{X}\right)^{2n}.
\end{eqnarray}
The log-term was evaluated above. The sum term in Eq.~(\ref{eq:line2}) should be evaluated to the leading order in $v\Lambda \Delta^{}_1\Delta^{}_2$. Explicitly it reads
\begin{eqnarray}
\sum^\infty_{n=1}q^{2n}\frac{2^{n-1}}{n}\frac{(2n-1)!!}{n!} \left(\frac{Y}{X}\right)^{2n} &=& \sum^\infty_{n=1}~q^{2n}\frac{2^{n-1}}{n}\frac{(2n-1)!!}{n!} 
\frac{(v\Delta^{}_1\Delta^{}_2)^{2n}}{[-iq^{}_0 (q^2_0+\Delta^2_1+\Delta^2_2) + \mu q^2_0 ]^{2n}}.
\end{eqnarray}
The momentum integration is easily performed 
\begin{eqnarray}
&& \int\frac{d^3Q}{(2\pi)^3}~ \sum^\infty_{n=1}~q^{2n}\frac{2^{n-1}}{n}\frac{(2n-1)!!}{n!} \frac{(v\Delta^{}_1\Delta^{}_2)^{2n}}{[-iq^{}_0 (q^2_0+\Delta^2_1+\Delta^2_2) + \mu q^2_0 ]^{2n}} \\
&=& \frac{\Lambda^2}{4\pi} \intop^\infty_{-\infty}\frac{dq^{}_0}{2\pi}~ \sum^\infty_{n=1}~\frac{2^{n-1}}{n}\frac{(2n-1)!!}{n!} 
\frac{(v\Lambda \Delta^{}_1\Delta^{}_2)^{2n}}{[-iq^{}_0 (q^2_0+\Delta^2_1+\Delta^2_2) + \mu q^2_0 ]^{2n}}.
\end{eqnarray}
The remaining frequency integral is difficult because every term in the series diverges for $q^{}_0=0$. On the other hand, putting $q^{}_0=0$ in Eq.~(\ref{eq:log}) does not 
suggest any singularity, because it is cut off by the momentum term. Therefore we may expect that the $q^{}_0\to0$-divergence of the frequency integral will disappear 
if all divergent terms are summed over. We rewrite the frequency integral as 
\begin{equation}
\intop^\infty_{-\infty}\frac{dq^{}_0}{2\pi} ~ \frac{1}{[-iq^{}_0 (q^2_0+\Delta^2) + \mu q^2_0]^{2n}}
= 
2\intop^\infty_{\lambda}\frac{dq^{}_0}{2\pi} ~ \frac{{\rm Re}[iq^{}_0 (q^2_0+\Delta^2) + \mu q^2_0]^{2n}}{[\mu^2q^4_0 + q^2_0(q^2_0+\Delta^2)^2]^{2n}},
\end{equation}
where we introduced the infrared cutoff $\lambda$ which will be sent to zero at the end of calculations. The imaginary part of the numerator is an odd function 
of the frequency and therefore disappears from the integral by symmetry. The contribution from the integral at the lower integration boundary consists of divergent terms, 
while the regular ones all disappear. The divergent terms are 
\begin{eqnarray}
n = 1: \frac{1}{\lambda \Delta^4}, \,\;\, n = 2: -\frac{1}{3\lambda^3 \Delta^8} + \cdots, \,\;\,
n = 3:  \frac{1}{5\lambda^5\Delta^{12}} -\cdots, \,\;\,n = 4:  -\frac{1}{7\lambda^7 \Delta^{16}} + \cdots, \; {\rm etc}
\end{eqnarray}
Hence we recognize an involved hierarchy of alternating series in odd inverse powers of $\lambda$, each starting at each $n$. The most divergent, which starts at $n=1$ has elements 
\begin{equation}
\label{eq:prima}
 \frac{1}{\lambda \Delta^4},\;\;  -\frac{1}{3\lambda^3 \Delta^8}, \;\;  \frac{1}{5\lambda^5\Delta^{12}}, \;\; -\frac{1}{7\lambda^7 \Delta^{16}}, \cdots \;\;  \lambda\frac{(-1)^{n-1}}{2n-1}\frac{1}{(\lambda \Delta^2)^{2n}}, \; n\geqslant1 .
\end{equation}
The alternating sign hints to the possible convergence.  The sequence gives rise to an infinite series which can be summed over separately giving the leading order term in powers of $v$:
\begin{eqnarray}
\frac{\Lambda^2}{(2\pi)^2}\frac{v\Lambda \Delta^{}_1\Delta^{}_2}{\Delta^2}
\lim_{\lambda\to0}\left(\frac{\lambda \Delta^2}{v\Lambda \Delta^{}_1\Delta^{}_2}\right)
\sum^\infty_{n=1}(-1)^{n-1} \frac{2^{n-1}}{n(n+1)!}\frac{(2n-1)!!}{2n-1}\left(\frac{v\Lambda \Delta^{}_1\Delta^{}_2}{\lambda \Delta^2}\right)^{2n}.
\end{eqnarray}
MATHEMATICA finds the sum to be a generalized hyper-geometric function
\begin{equation}
 \sum^\infty_{n=1}(-1)^{n-1} \frac{2^{n-1}}{n(n+1)!}\frac{(2n-1)!!}{2n-1} x^{2n} = \frac{x^2}{2}~ {}^{}_pF^{}_q\left(\left\{\frac{1}{2},1,1\right\},\left\{2,3\right\},-4x^2\right),
\end{equation}
which grows linearly for large $x$ (i.e. small $\lambda$). Therefore, the limit $\lambda\to0$ exists:
\begin{equation}
 \lim_{x\to\infty}\frac{x}{2}~ {}^{}_pF^{}_q\left(\left\{\frac{1}{2},1,1\right\},\left\{2,3\right\},-4x^2\right) = \frac{4}{3}.
\end{equation}
Hence, the leading correction from the lower integral boundary (i.e. with a global minus sing) is linear in mixed terms of $\Delta^{}_2\Delta^{}_2$:
\begin{eqnarray}
&&-\int\frac{d^3Q}{(2\pi)^3}~ \sum^\infty_{n=1}~q^{2n}\frac{2^{n-1}}{n}\frac{(2n-1)!!}{n!} \frac{(v\Lambda\Delta^{}_1\Delta^{}_2)^{2n}}{[-iq^{}_0 (q^2_0+\Delta^2_1+\Delta^2_2) + \mu q^2_0 ]^{2n}} \\
&&\approx \frac{4}{3}\frac{\Lambda^2}{(2\pi)^2}\frac{v\Lambda \Delta^{}_1 \Delta^{}_2}{\Delta^2_1+\Delta^2_2} - {\rm cotribution\, from\, the \, upper\, cutoff\,}q^{}_0\to\infty. 
\end{eqnarray}
The contribution from the upper cutoff is not important, since it is at least quadratic in powers of $v$ and therefore subdominant to the just obtained term. 
The free energy functional to this order and zero Dirac mass reads

\begin{equation}
{\cal F}^{(2)}_{MF} \approx \frac{\mu\Lambda^2}{2\sqrt{2}\pi}
\left[ 
\frac{\bar\Delta^2_1+\bar\Delta^2_2}{\gamma} - \left(\epsilon^{\frac{1}{2}}_+ + \epsilon^{\frac{1}{2}}_- + 
\frac{4\bar v}{3\sqrt{2}\pi}\frac{\bar\Delta^{}_1\bar\Delta^{}_2}{\bar\Delta^2_1+\bar\Delta^2_2} \right)
\right],
\end{equation}
where $\bar v = v\Lambda/\mu$, $\bar\Delta^{}_1 = \Delta^{}_1/\mu$, $\bar\Delta^{}_2 = \Delta^{}_2/\mu$, $\displaystyle\gamma = \frac{g\Lambda^2}{\sqrt{2}\pi\mu}$, and
\begin{equation}
\label{eq:spr12}
\epsilon^{}_\pm = 1+2 \bar\Delta^2_1+2\bar\Delta^2_2\pm \sqrt{1+4 \bar\Delta^2_1+4\bar\Delta^2_2}.
\end{equation}
In units of ${2\sqrt{2}\pi}/{\mu\Lambda^2}$ all evaluated contributions to the effective potential become
\begin{eqnarray}
\label{eq:MFEF}
\tilde{\cal F}^{}_{MF} & \approx &
\frac{\bar\Delta^2_1+\bar\Delta^2_2}{\gamma} - 
\left(\epsilon^{\frac{1}{2}}_+ + \epsilon^{\frac{1}{2}}_- + 
\frac{\bar v\bar\Delta^{}_1\bar\Delta^{}_2}{\bar\Delta^2_1+\bar\Delta^2_2}  
+
 \frac{\bar\Delta^{}_{\rm D}(\bar\Delta^2_1-\bar\Delta^2_2)}
{\sqrt{1+4(\bar\Delta^2_1+\bar\Delta^2_2)}}
\frac{\epsilon^{\frac{1}{2}}_+ - \epsilon^{\frac{1}{2}}_-}{\epsilon^{\frac{1}{2}}_- \epsilon^{\frac{1}{2}}_+}
\right).
\end{eqnarray}
We plot this expression in the main part in Fig.~\ref{fig:Fig3}. The gradient flow shown in Fig.~\ref{fig:Fig4}  is calculated from 
\begin{equation}
\left\{
-\frac{\delta}{\delta\bar\Delta^{}_1}\tilde{\cal F}^{}_{MF} , -\frac{\delta}{\delta\bar\Delta^{}_2}\tilde{\cal F}^{}_{MF}
\right\}.
\end{equation}

\subsection{4. The spectrum of the Hamiltonian (\ref{eq:MFH})}

\subsection{4.1 Cardano solutions of the secular equation}
\label{app:cardano}

\begin{figure}[t]
\includegraphics[width=5.75cm]{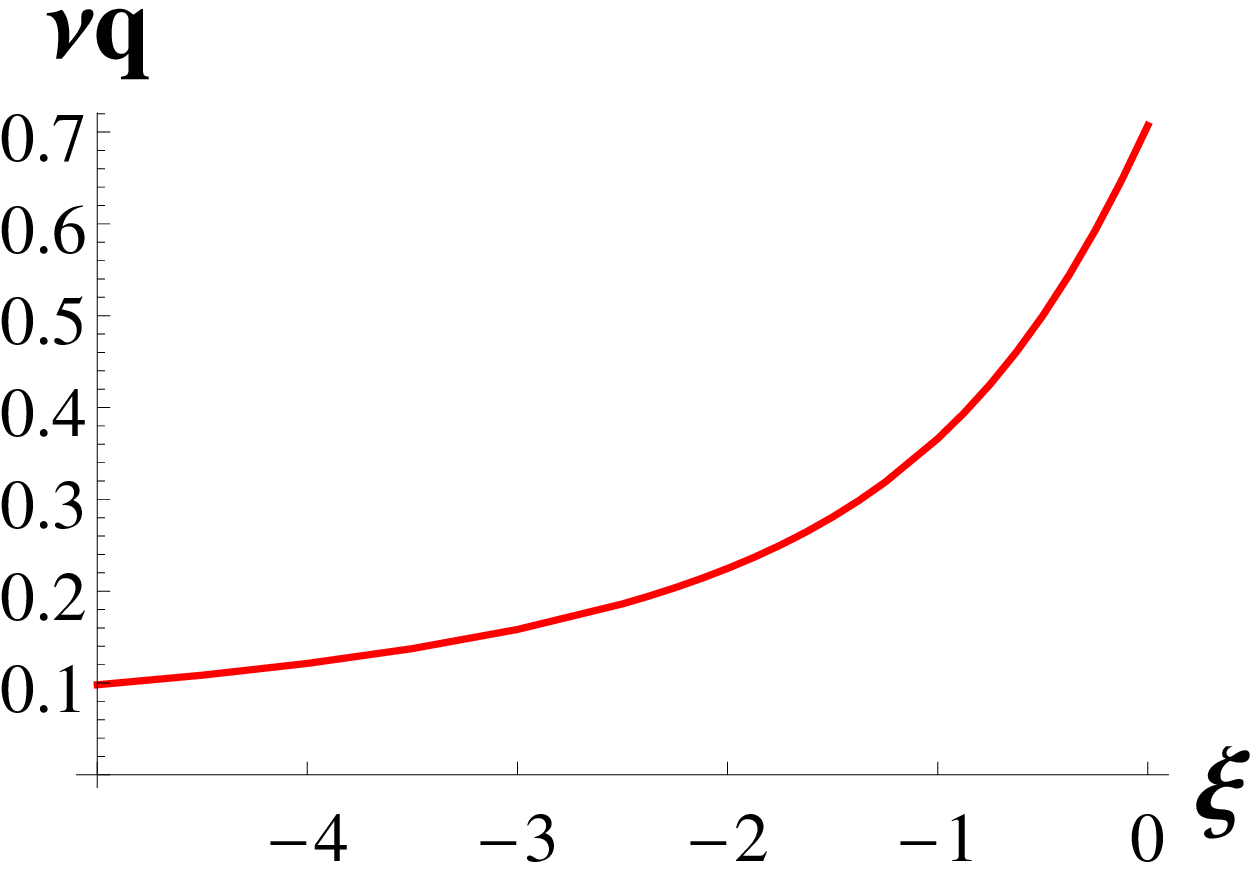}
\includegraphics[width=5.75cm]{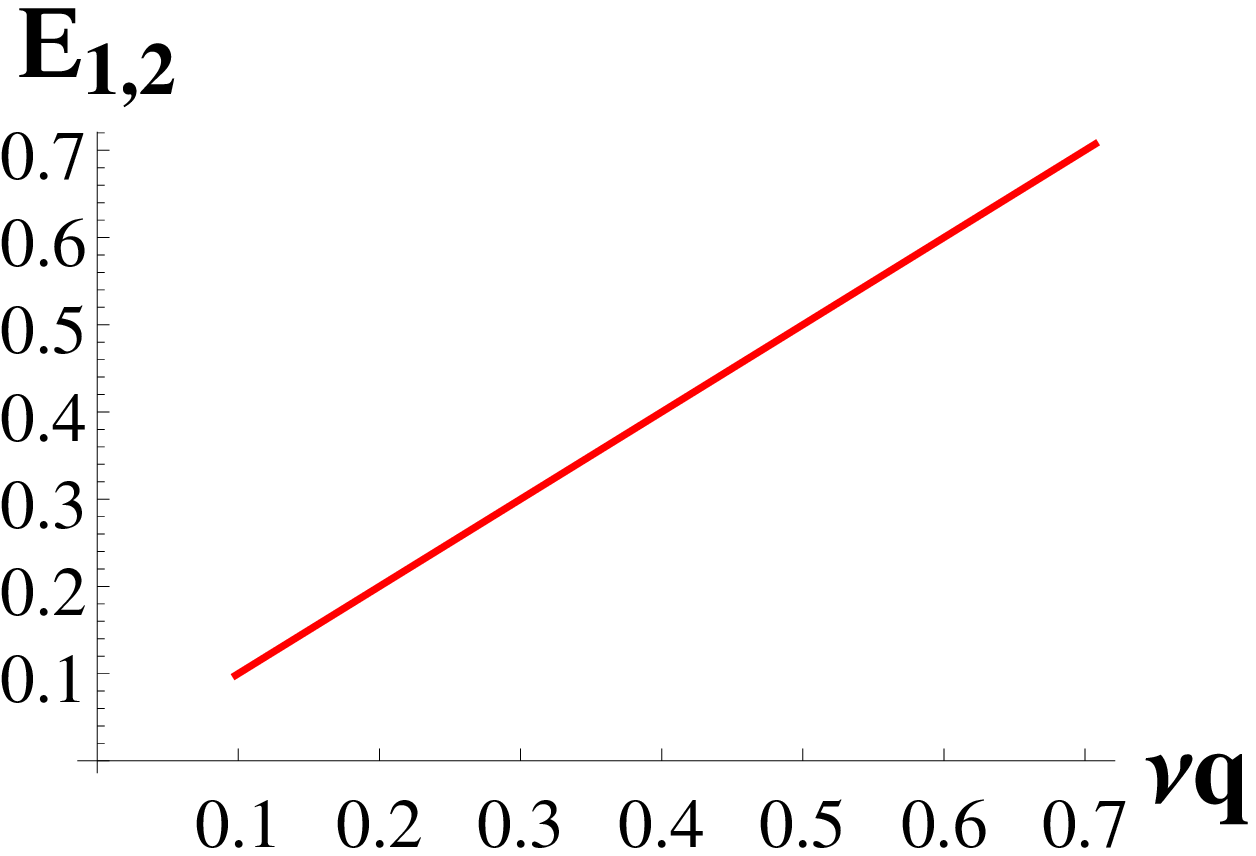}
\caption{ 
The parametric line of the (lower) touching point formation (left) and eigenvalues $E^{}_2$ and $E^{}_3$ at the touching point (right) as function of the momentum. 
The energy scales in units of $\Delta$.
}
\label{fig:FigS2}
\end{figure}

The eigenvalues $E^{}_i$ of the mean-field Hamiltonian (\ref{eq:MFH}) for zero Dirac mass are found from the secular equation 
\begin{equation}
\label{eq:sec}
\left|
\begin{array}{ccc}
          - E               &   vq e^{i\phi}        &  \Delta^{}_1 e^{i\chi^{}_1}  \\  
   vq e^{-i\phi}     &          - E                &   \Delta^{}_2 e^{i\chi^{}_2} \\
 \Delta^{}_1e^{-i\chi^{}_1} & \Delta^{}_2 e^{-i\chi^{}_2} &  \xi^{}_q  - E         
\end{array}
\right|=0,
\end{equation}
which becomes 
\begin{equation}
\label{eq:SecEq}
-E^3+a E^2+bE+c=0, 
\end{equation}
with 
\begin{equation}
a = \xi^{}_q, \;\; b = \Delta^2_1 + \Delta^2_2 + v^2q^2,\;\; c = 2vq\Delta^{}_1\Delta^{}_2 \cos\left(\chi^{}_{2} - \chi^{}_{1} + \phi\right) - v^2q^2\xi^{}_q. 
\end{equation}
The solutions of the secular equation are given by the Cardano formulas:
\begin{eqnarray}
\label{eq:Cord1}
E^{}_1 &=& \frac{1}{3}\left[a - (a^2+3b)\left(\frac{2}{A}\right)^{\frac{1}{3}} - \left(\frac{A}{2}\right)^{\frac{1}{3}} \right], \\
\label{eq:Cord2}
E^{}_2 &=& \frac{1}{3}\left[a + e^{i\frac{\pi}{3}}(a^2+3b) \left(\frac{2}{A}\right)^{\frac{1}{3}}  + e^{-i\frac{\pi}{3}}  \left(\frac{A}{2}\right)^{\frac{1}{3}} \right], \\
\label{eq:Cord3}
E^{}_3 &=& \frac{1}{3}\left[a + e^{-i\frac{\pi}{3}}(a^2+3b) \left(\frac{2}{A}\right)^{\frac{1}{3}} + e^{i\frac{\pi}{3}}  \left(\frac{A}{2}\right)^{\frac{1}{3}} \right],
\end{eqnarray}
where 
\begin{equation}
A = \left( 3\sqrt{3}i\sqrt{a^2b^2+4b^3-4a^3c-18abc-27c^2} - 2a^3 - 9ab - 27c \right).
\end{equation}
The three eigenvalues do fulfill the usual constrain conditions:
\begin{eqnarray}
\displaystyle\sum^3_{i=1}E^{}_i  = {\rm tr}[{\rm H}^{}_{\rm MF}] = \xi^{}_q, && 
\displaystyle\prod^3_{i=1}E^{}_i = {\rm det}[{\rm H}^{}_{\rm MF}] = 2vq\Delta^{}_2\Delta^{}_1 \cos\left(\chi^{}_{2} - \chi^{}_{1} + \phi\right) - v^2q^2\xi^{}_q.
\end{eqnarray}
The form of modes $E^{}_2$ and $E^{}_3$ suggests that they may become equal if 
\begin{equation}
 (a^2+3b) \left(\frac{2}{A}\right)^{\frac{1}{3}} = \left(\frac{A}{2}\right)^{\frac{1}{3}},
\end{equation}
which occurs for special values of $vq$ and $\xi^{}_q$. The plot of this numerically evaluated curve at the energetically lower touching point is shown in Fig.~\ref{fig:FigS2}.

\subsection{4.2 Crossing point condition for two upper bands}
\label{app:factorization}

\begin{figure}[t]
\includegraphics[width=5.2cm]{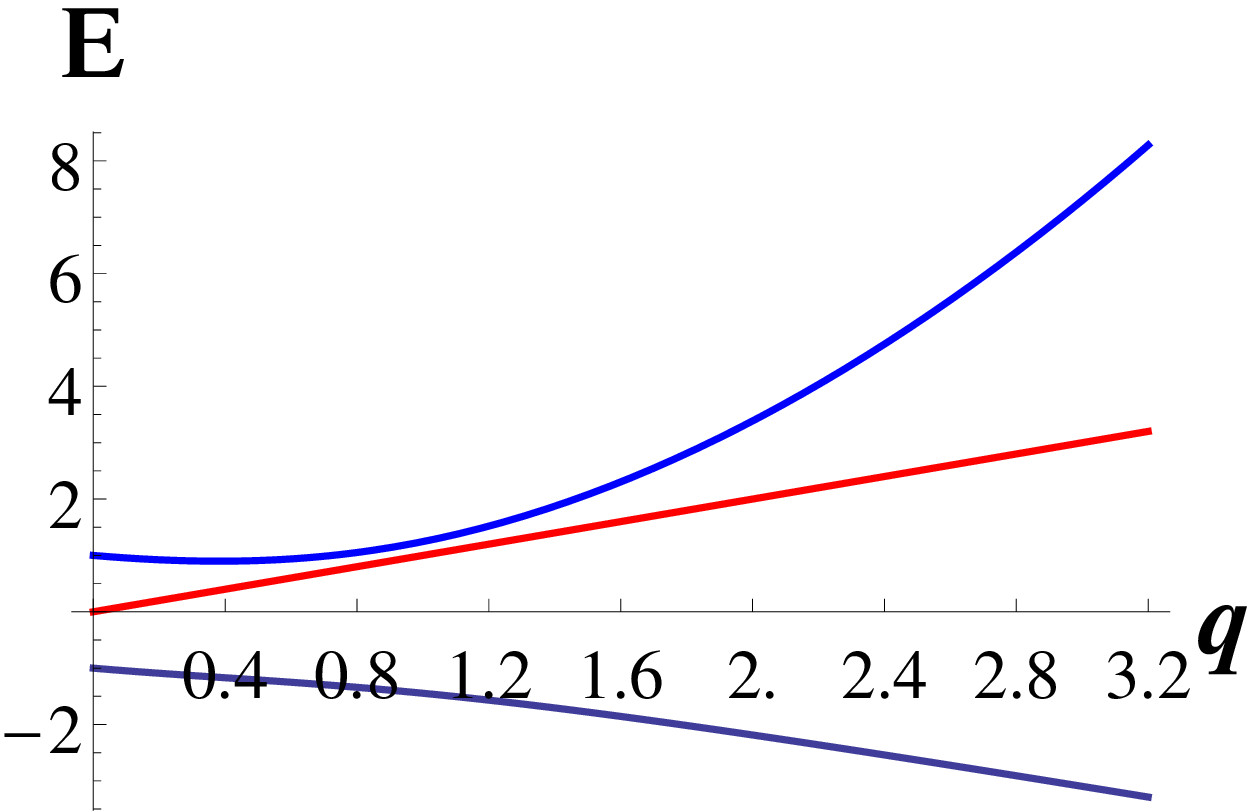}
\includegraphics[width=5.2cm]{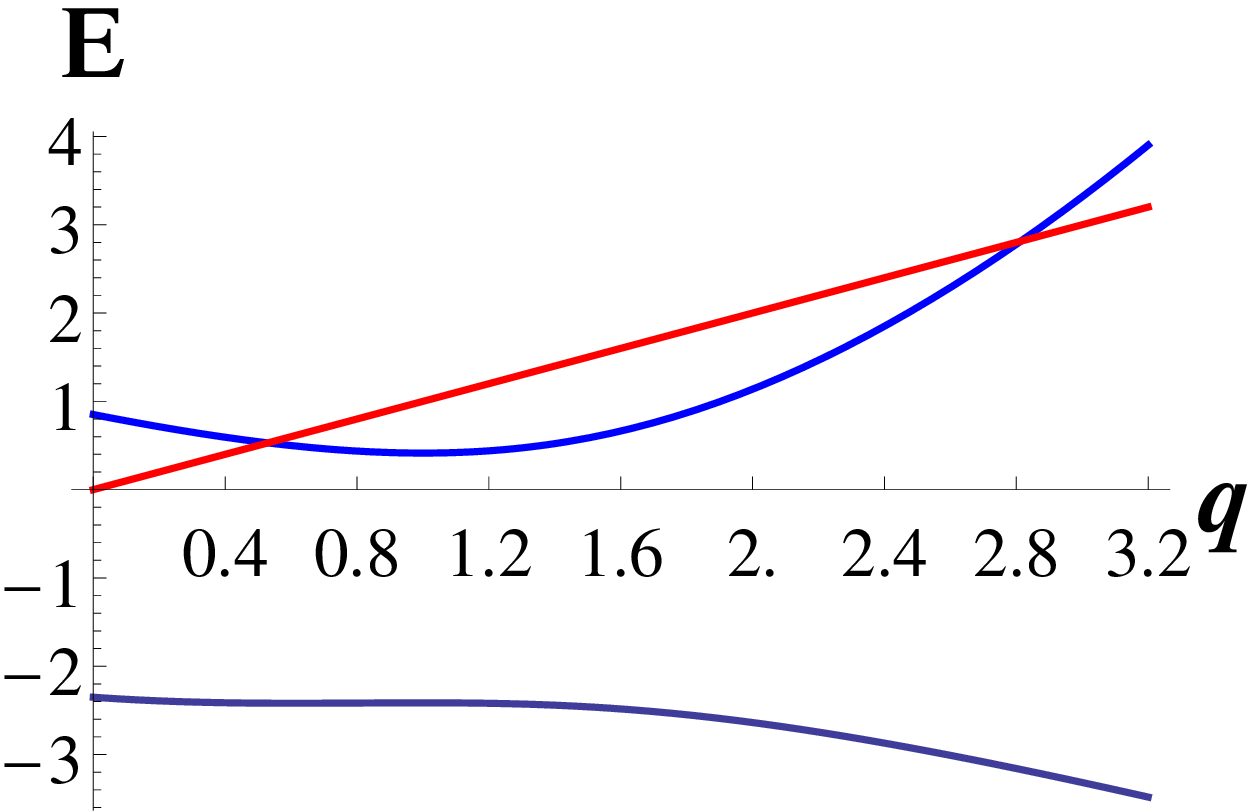}
\includegraphics[width=5.2cm]{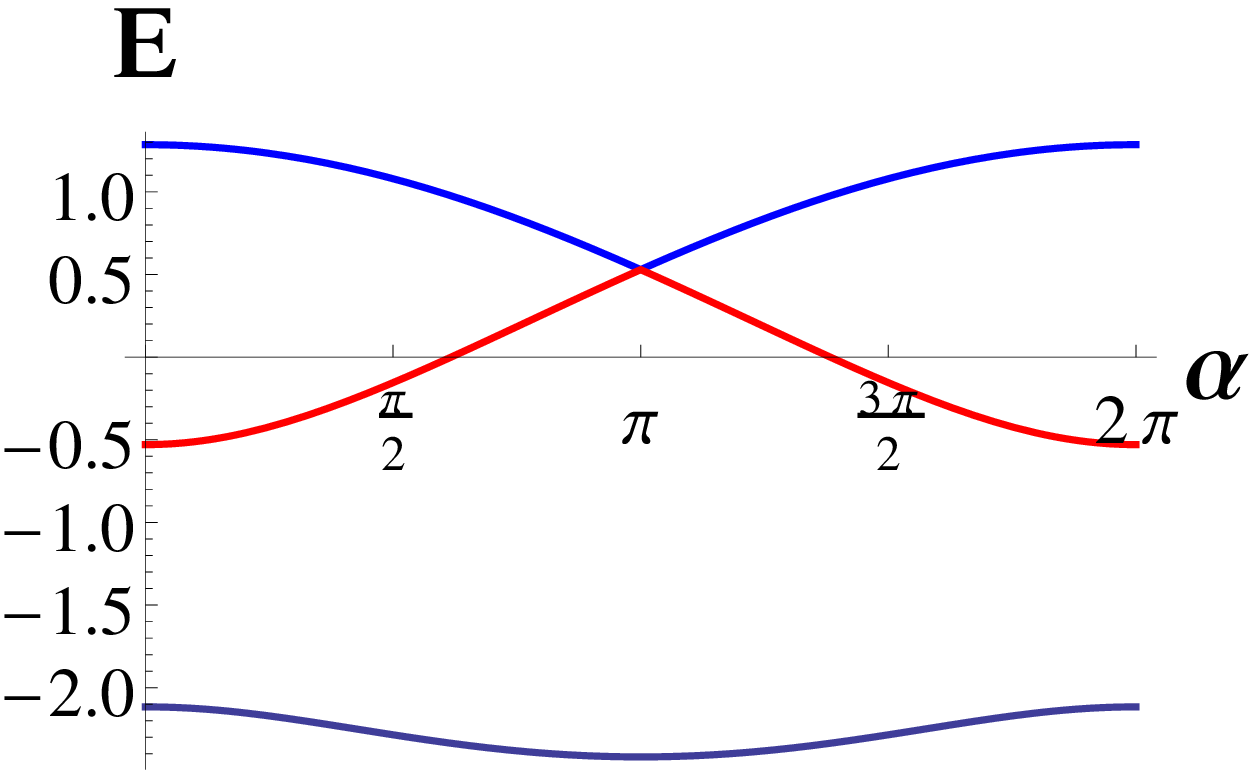}
\caption{Here we use the notation $\alpha=\phi+\chi^{}_{2} - \chi^{}_{1}$. 
Left panel: Dispersion of all spectral branches at $\alpha=\pi$ in the parametric regime without crossing points.
Middle panel: Dispersion of all spectral branches at $\alpha=\pi$ in the parametric regime with crossing points as function of momentum. 
Right panel: All spectral branches taken at the crossing point at lower energy as function of the polar angle $\alpha$.
Middle and right panel panel are two different projections of the same curve and are plotted with the same parameter values as
for right figure in Fig.~(\ref{fig:Fig5}). On the shape of the lowest band one recognizes the saddle point responsible for the sharp peak 
in the DOS from Fig.~(\ref{fig:Fig7}). Its rough coordinates are $\alpha=\pi$, $q\approx 1.2$.
} 
\label{fig:FigS3}
\end{figure}

\begin{figure}[t]
\includegraphics[width=5.75cm]{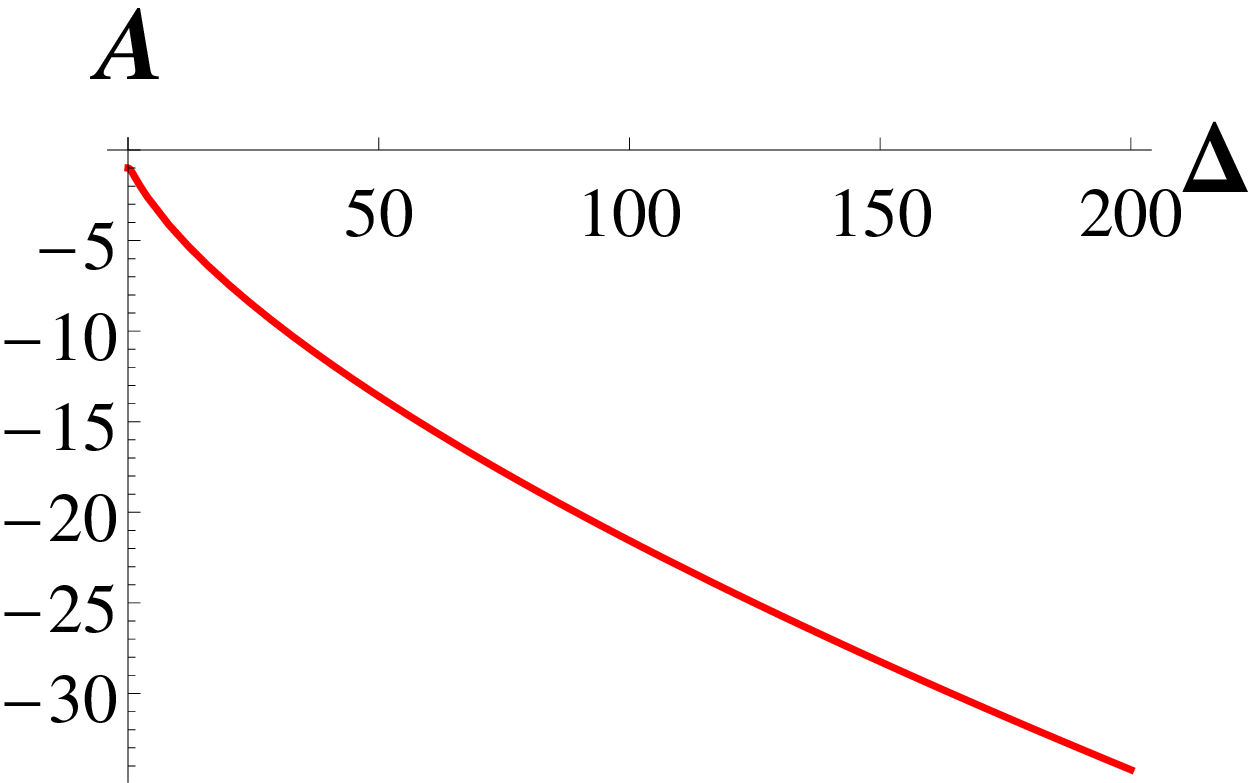}
\hspace{4mm}
\includegraphics[width=5.75cm]{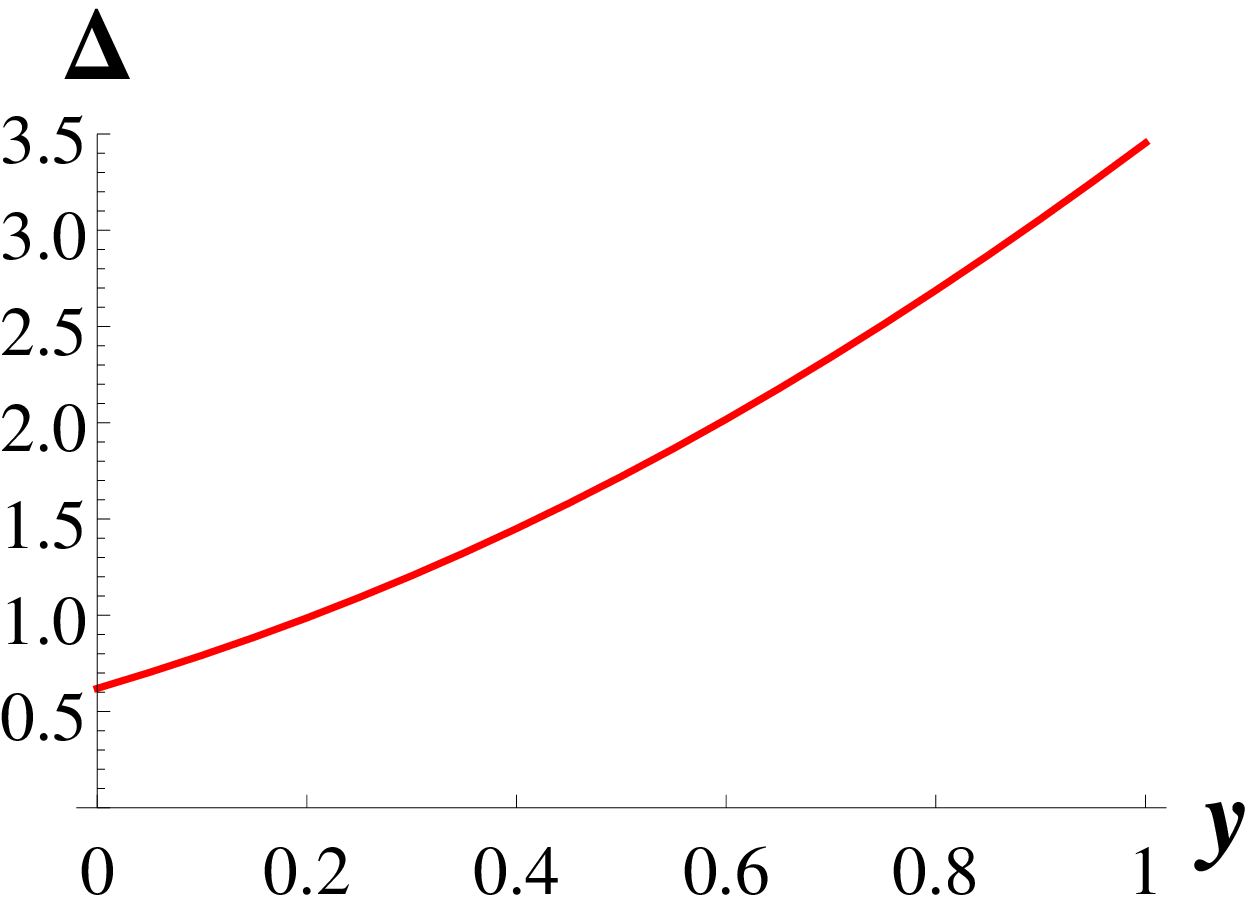}
\caption{ 
Left panel: The factor $A$ from Eq.~(\ref{eq:AssDet3}) as function of the dimensionless order parameter $\bar\Delta$.
Right panel: Numerical evaluation of the criterion Eq.~(\ref{eq:Crit}) for the dimensionless order parameter $\bar\Delta$ as function of 
the parameter $y=mv/q^{}_F$. 
}
\label{fig:FigS4}
\end{figure}

Here we define the conditions for crossing points which appear between the two upper bands defined in Eq.~(\ref{eq:cr1}) and shown in Fig.~\ref{fig:FigS3}.
If we fix the phase to $\phi=\pi$ (i.e. $q=|q^{}_x|$), put $\Delta^{}_{\rm D}=0$ and utilize the variational equation solution 
$\Delta^{}_1=\Delta/\sqrt{2}=\Delta^{}_2$ and $\chi^{}_{2} = \chi = \chi^{}_{1}$, 
then the determinant of the mean-field Hamiltonian Eq.~(\ref{eq:MFH}) reduces to 
\begin{equation}
\label{eq:AssDet0}
\det[{\rm H}^{\rm }_{\rm MF}] = -vq(vq\xi^{}_q+\Delta^2),
\end{equation}
which is the product of all three eigenvalues shown in Eq.~(\ref{eq:cr1}). 
Introducing the dimensionless momenta $\bar q=q/q^{}_F$ and the order parameter $\bar\Delta^2=\Delta^2/(vq^3_F/2m)$ 
we rewrite Eq.~(\ref{eq:AssDet0}) as
\begin{equation}
\label{eq:AssDet1}
\det[{\rm H}^{\rm }_{\rm MF}] = -\frac{v^2 q^4_F}{2m}\bar q(\bar q^3 - \bar q + \bar\Delta^2). 
\end{equation}
We make a decomposition guess dictated by the expected asymptotics of the eigenvalues
\begin{equation}
\label{eq:AssDet2}
\det[{\rm H}^{\rm }_{\rm MF}] = -\frac{v^2 q^4_F}{2m}\bar q (\bar q^2 + A\bar q + B)(\bar q+C),
\end{equation}
which yields by comparison with Eq.~(\ref{eq:AssDet1}) the following relations for the decomposition coefficients
\begin{equation}
 A  + C =  0, \;\;\; 
 AC + B = -1, \;\;\;
BC = \bar\Delta^2,  
\end{equation}
which furthermore leads to the solutions
\begin{equation}
C = -A,\;\;\;  B = -\frac{\bar\Delta^2}{A}, 
\end{equation}
and an equation for $A$
\begin{equation}
\label{eq:AA}
A^3-A+\bar\Delta^2 = 0. 
\end{equation}
There is only one real solution for $A$ 
\begin{equation}
\label{eq:AssDet3}
A =  
\frac{\displaystyle -2\cdot3^{\frac{1}{3}}e^{i\frac{\pi}{3}} - 2^{\frac{1}{3}}e^{-i\frac{\pi}{3}}\left(\sqrt{81\bar\Delta^4-12} - 9\bar\Delta^2\right)^{\frac{2}{3}}  }
{\displaystyle 6^{\frac{2}{3}} \left(\sqrt{81\bar\Delta^4-12} - 9\bar\Delta^2\right)^{\frac{1}{3}}},
\end{equation}
which is shown in Fig.~(\ref{fig:FigS4}). The factorization Eq.~(\ref{eq:AssDet0}) becomes explicitly 
\begin{equation}
\label{eq:AssDet4}
\det[{\rm H}^{\rm }_{\rm MF}] = -vq^{}_F(\bar q-A)\cdot\frac{q^2_F}{2m}\left(\bar q^2 + A\bar q - \frac{\bar\Delta^2}{A}\right)\cdot vq^{}_F\bar q,
\end{equation}
which allows us to determine the position of the touching points in the momentum space from 
\begin{equation}
\frac{q^2_F}{2m}\left(\bar q^2 + A\bar q - \frac{\bar\Delta^2}{A}\right) = vq^{}_F\bar q,
\end{equation}
which leads to the quadratic equation  
\begin{equation}
\bar q^2 +\left(A-\frac{2mv}{q^{}_F}\right) \bar q - \frac{\bar\Delta^2}{A} = 0, 
\end{equation}
which has the solutions 
\begin{equation}
\label{eq:coord}
\bar q^{}_\pm = \frac{1}{2}\left[
\frac{2mv}{q^{}_F} - A \pm \sqrt{\left(A - \frac{2mv}{q^{}_F}\right)^2 + 4 \frac{\bar\Delta^2}{A}}
\right] ,
\end{equation}
$q^{}_-$ corresponding to the crossing point at smaller and $q^{}_+$ at larger energies. 
Taken into account that $A$ is actually negative, cf. Fig.~\ref{fig:FigS4}, the crossing points form only if 
\begin{equation}
\label{eq:Crit}
\left(|A| + \frac{2mv}{q^{}_F}\right)^2 \geqslant 4 \frac{\bar\Delta^2}{|A|},
\end{equation}
from which the condition for the  critical value of the pairing order parameter $\Delta$ can be found. 
The strict equality in Eq.~(\ref{eq:Crit}) defines the critical order parameter for the given system realization at which the gap between the two bands closes.
The numerically evaluated order parameter as function of the ratio $y=mv/q^{}_F$  is shown in Fig.~\ref{fig:FigS4}.

\end{document}